\documentclass[10pt]{iopart}

\usepackage[T1]{fontenc}
\usepackage[utf8]{inputenc} 
\usepackage[english]{babel}

\usepackage{verbatim}
\usepackage{caption}

\usepackage{wrapfig}

\usepackage{subfigure}

\usepackage{iopams}
\usepackage{hyperref}

\expandafter\let\csname equation*\endcsname\relax

\expandafter\let\csname endequation*\endcsname\relax

\usepackage{amsmath}

\usepackage{pgfplots}
\pgfplotsset{compat = newest}
\usepackage{pgfplotstable}
\pgfplotsset{ every non boxed x axis/.append style={x axis line style=-},
     every non boxed y axis/.append style={y axis line style=-}}
\usetikzlibrary{pgfplots.groupplots}

\usepackage{algorithm}
\usepackage{algpseudocode}
\algtext*{EndIf}
\newcommand{\algorithmicoutput}{\textbf{Output:}}
\newcommand{\Output}{\item[\algorithmicoutput]}

\usepackage{graphicx, floatflt}

\usepackage{tikz}
\usetikzlibrary{plotmarks}

\begin{document}

\title{Adaptive Cluster Expansion for Ising spin models}
\author{Simona Cocco$^1$, Giancarlo Croce$^2$, Francesco Zamponi$^1$
}
\address{$^1$ Laboratoire de Physique de l'Ecole Normale Sup\'erieure, ENS, Universit\'e PSL, CNRS, Sorbonne Universit\'e, Universit\'e de Paris, Paris, France}
\address{$^2$  Sorbonne Universit\'e, CNRS, Institut de Biologie Paris Seine, Biologie computationnelle et quantitative – LCQB, 75005 Paris, France}

\date{}

\begin{abstract}
We propose an algorithm to obtain numerically approximate solutions of
the {\em direct} Ising  problem,
 that is, to compute the free energy and the equilibrium observables of 
 spin systems with arbitrary two-spin interactions.
 To this purpose we use the Adaptive Cluster Expansion method~\cite{Monasson-Cocco}, originally developed to solve the {\em inverse} Ising problem, 
 that is, to infer the interactions from the equilibrium correlations.
The method consists in iteratively  constructing and selecting
clusters of spins,  computing their contributions to the free energy and discarding clusters whose contribution is lower than a fixed threshold. The properties of the cluster expansion and its performance are studied in detail on  one dimensional, two  dimensional, random  and fully connected graphs with homogeneous or heterogeneous fields and couplings. 
We discuss the differences between different representations (Boolean and Ising) of the spin variables.
\end{abstract}

\tableofcontents

\newpage

\section{Introduction}

We consider a finite number $N$ of binary variables, either Boolean variables
 $\sigma_i \in \lbrace 0,1 \rbrace $ or Ising spin variables $s_i \in \lbrace -1,1 \rbrace$, interacting via a set of $N$ local fields $h_i$ and of $\frac{1}{2} N (N-1)$ pairwise couplings $J_{ij}$. We define  $\Omega \doteq \lbrace 1, \ldots, N \rbrace$ and $\mathcal{J} = \lbrace \lbrace h_i \rbrace, \lbrace J_{ij} \rbrace \rbrace$. 
The energy of a configuration $\boldsymbol{\sigma}=\lbrace \sigma_1, \ldots, \sigma_N \rbrace  \in \lbrace 0,1 \rbrace ^N$ 
is given by the Hamiltonian:
\begin{equation}\label{eq:H}
H_{\mathcal{J} } ( \boldsymbol{\sigma} )= - \sum_i {h_i} \sigma_i - \sum_{i<j} {J_{ij}} \sigma_i \sigma_j.
\end{equation}
In statistical mechanics, one is usually interested in computing the expectation values of certain physical quantities, like 
the free energy of the system in thermal equilibrium: 
\begin{equation} \label{freeE}	
	F_{\mathcal{J}} = -\beta^{-1} \log Z({\mathcal{J}}) \ ,
\end{equation} 
and the set $\mathcal{C}$ of the equilibrium correlations, meaning the average values of the spins and 
the spin-spin correlations over the Gibbs measure corresponding to the Hamiltonian \eqref{eq:H}:
\begin{equation} \label{magne}	
\mathcal{C} = \lbrace \lbrace \langle \sigma_i \rangle \rbrace, \lbrace \langle \sigma_i \sigma_j \rangle \rbrace \rbrace \ .
\end{equation} 
Both calculations require the computation of the partition function
\begin{equation} 	
	Z({\mathcal{J}}) = \sum_{\boldsymbol{\sigma}} \e^{-\beta H_{\mathcal{J} } ( \boldsymbol{\sigma} )} \ ,
\end{equation} 
the required computational time being $\mathcal{O}(2^N)$.
Practically speaking, for a large system the calculation then becomes impossible,
and approximation methods are required.

This is the problem we are interested in this paper, and we refer to it as the {\it direct Ising problem}.
The {\it inverse Ising problem} consists, on the contrary, in finding the interaction set $\mathcal{J}$ 
from the measure of the correlations $\mathcal{C}$ \cite{Monasson-Cocco,Barton:2016:ACE,nguyeninv17,CoccoInv}. 
The \textit{Adaptive Cluster Expansion} (ACE) method~\cite{Monasson-Cocco,Monasson-Cocco2} has been originally developed to solve the inverse problem.
The aim of this paper is to adapt the ACE method to solve the direct
Ising problem: in particular, we want to compute the free energy and the correlations $\mathcal{C}$ of the model from the set $\mathcal{J}$.

Let us stress from the very beginning that, of course, several very efficient methods to solve the direct Ising problem are available in the literature. 
In the following, and in the bulk of the paper,
we review them shortly and discuss what kind of advantages (and disadvantages) the ACE method could offer with respect to those methods.
As a general comment, we believe that the ACE method has the advantage of being very general and simple to implement; but on the other hand,
it is certainly not competitive for large-scale simulations of ordered systems. Its efficiency for disordered systems should be discussed on a case-to-case
basis. If one is interested in obtained a good (but not necessarily excellent) approximation of the free energy and correlations for a large ensemble of
rather small systems with generic random couplings and fields, then we believe that the ACE method offers a good alternative to existing methods and
could outperform Monte Carlo schemes if disorder is strong enough. Note that the situation described above is quite common in applications to biological
problems, see e.g.~\cite{rivoire}.

Several cluster methods have been used before to solve the direct Ising problem. 
Indeed, for ordered models defined on a regular lattice, the set of clusters that should be kept in the expansion 
can be chosen ad hoc by following the symmetries of the problem, as in the Ising model~\cite{rew1_EPJB_a}. 
In fully connected model or large-dimensional models, both ordered and disordered, only a finite number of clusters contribute in the thermodynamic limit~\cite{yedidia2}; 
examples are the Sherrington-Kirkpatrick model~\cite{rew1_EPJB_b,thouless1977} or even more general spin systems such as the Heisenberg model~\cite{rew1_EPJB_c}. 
However, these methods cannot be easily extended to general disordered systems where underlying symmetries are missing and all clusters are potentially relevant. 
The ACE algorithm, on the contrary, is designed to automatically find the most relevant clusters.
Other expansion schemes such as belief propagation can provide an estimation of the free energy, at least for sparse graphs with few loops~\cite{yedidia}.
 Generalisations of belief propagation \cite{yedidia,Opper-Saad} and cluster expansions~\cite{kikuchi96,yedidia2,an88,pelizzola05,yasuda06} have been used to construct several more sophisticated expansion methods.
Compared to these classical cluster expansions, the ACE method is simpler to implement, because it does not require any consistency equation 
between the marginals calculated from different clusters, or between a cluster and the whole system.

A major problem in cluster expansion methods is how to choose clusters and how to sum them up in an appropriate way to obtain the partition function. 
Here, we base our cluster expansion on a construction rule which selectively builds up larger clusters from smallest ones,
 and on a selection rule to sum up clusters according to the amplitudes of their free energy contributions~\cite{Monasson-Cocco,Monasson-Cocco2}. 
 It is important to point out that the method is not exact:
for instance, because we do not impose any consistency equation, the choice of the representations of the variables (e.g. Ising or Boolean) 
is very important for the rapid convergency of the algorithm, as we will discuss in the following. 
Nevertheless, we find that the method has good performance in interesting cases.

The Monte Carlo (MC) technique is also very efficient to compute the spin correlations, but the calculation of the free energy
via Monte Carlo is less efficient, because it requires a thermodynamic integration~\cite{Free_Energy,d,frenkel}.
Several biased Monte Carlo sampling schemes have been developed in this context to better evaluate the free energy of the model.
An exhaustive review of these methods goes beyond the scope of the paper;
we mention here as examples 
importance sampling~\cite{d}, the Wang-Landau approach~\cite{wanglandau}, the component-distribution approach~\cite{hartmann},
cluster methods~\cite{d,houdayer}, umbrella sampling~\cite{d,frenkel} and tethered methods~\cite{bea}.
In the following, we will limit ourselves to compare our ACE implementation with the simplest MC scheme, namely 
a standard MC using thermodynamic integration.

In the following, we first present the ACE method (section~\ref{sec:cluster-exp}), and
then apply it to some classical examples of Ising models (sections ~\ref{sec:examples-ord} and \ref{sec:examples-disord}), both on ordered and disordered systems and at different temperatures.
We conclude by some comments on the comparison with Monte Carlo performances, and perspectives for future applications of the method
(section~\ref{sec:conclusion}).

\section{Cluster expansion} \label{sec:cluster-exp}

Let us call a {\textit{cluster}} $\Gamma$ a subset of $\Omega$ and consider a function $W_{\Gamma}(\mathcal{J}_{\Gamma})$
that depends only on the interaction parameters $\mathcal{J}_{\Gamma} = \lbrace h_i, J_{ij} \rbrace_{i,j \in {\Gamma}}$ that involve variables in 
the cluster $\Gamma$. 
We can derive the  contribution of the cluster $\Gamma$ to such function, $\Delta W_{\Gamma}$, from the following recursive formula:
\begin{equation}\label{clust_exp}
\begin{split} 
& \Delta W_{i} (\mathcal{J}_{i}) = W_{i} (\mathcal{J}_{i}) \ , \\
& \Delta W_{\Gamma} (\mathcal{J}_{\Gamma}) = W_{\Gamma} (\mathcal{J}_{\Gamma}) - 
\sum_{\Gamma' \subset \Gamma} \Delta W_{\Gamma'}(\mathcal{J}_{\Gamma'}) \ , \end{split}
\end{equation} 
and therefore:
\begin{equation} \label{clust_exp_final} 
W_{\Gamma} (\mathcal{J}_{\Gamma}) = \sum_{\Gamma' \subseteq \Gamma} \Delta W_{\Gamma'}(\mathcal{J}_{\Gamma'}) \ .
\end{equation} 
For the whole system we thus get:
\begin{equation} \label{clust_approx}
W_{\Omega} (\mathcal{J}) = \sum_{\Gamma \subseteq \Omega} \Delta W_\Gamma(\mathcal{J}_{\Gamma}) \ .
\end{equation} 
Eq.~(\ref{clust_approx}) is an exact expansion, called M\"obius transform,  and the total number of clusters $\Gamma$ that appear in the sum is: 
\begin{equation}
{N \choose 1} + {N \choose 2} + \ldots + {N \choose N} = \sum_{k=0}^{N} {N \choose k}  -{N \choose 0} = 2^N -1 \ .
\end{equation}
The expansion holds for an arbitrary function, but here we are interested in
\begin{equation}
W_{\Gamma} (\mathcal{J}_{\Gamma}) = \log Z_{\Gamma} (\mathcal{J}_{\Gamma}) = \log \sum_{\sigma_i, i \in \Gamma} e^{\sum_{i \in \Gamma} \beta h_i \sigma_i + \sum_{\lbrace i ,j \rbrace \subseteq \Gamma } \beta J_{ij} \sigma_i \sigma_j} \ ,
\end{equation}
in such a way that $W_{\Omega} (\mathcal{J})$ 
is the \textit{log partition function}. Here, $\beta$ is the inverse temperature and fixes the overall scale of the couplings. 

The free energy is simply related to the log partition function by $F_{\Omega}(\mathcal{J}) = - \beta^{-1} W_{\Omega} (\mathcal{J})$.
From the log partition function, differentiating with respect to $h_i$ and $J_{ij}$ respectively, we can compute the correlations in Eq.~(\ref{magne}). 
Because the functions $W_{\Gamma}$ and $\Delta W_{\Gamma}$ depend only on the couplings in the cluster $\Gamma$, we have:
\begin{equation}\label{eq:Corr}
\begin{split}
\langle \sigma_i \rangle &=\frac1\beta \frac{d W_{\Omega} (\mathcal{J})}{d h_i} =\frac1\beta \sum_{\lbrace i \rbrace \subseteq \Gamma  \subseteq \Omega} \frac{d \Delta W _{\Gamma} (\mathcal{J}_{\Gamma})}{d h_i} \ , \\
\langle \sigma_i \sigma_j \rangle &= \frac1\beta\frac{d W_{\Omega} (\mathcal{J})}{d J_{ij}} = \frac1\beta\sum_{\lbrace i,j \rbrace \subseteq \Gamma  \subseteq \Omega} \frac{d\, \Delta W _{\Gamma} (\mathcal{J}_{\Gamma})}{d J_{ij}}\ . 
\end{split}\end{equation}

The expansion in Eq.(\ref{clust_approx}) is useless if one has to compute all the clusters: the total number of clusters is $2^N-1$ and computing
large clusters is equivalent to a full computation of the partition function. We therefore need an accurate truncation scheme, that allows one to keep only
a certain number of small enough clusters and yet provides a good approximation for the free energy and the correlations.
First, we note that only connected clusters contribute to the expansion for the log partition function.
A cluster $\Gamma$ is {\it disconnected} if it can be separated
in two disjoint subsets, $\Gamma = \Gamma_1 \cup \Gamma_2$ with $\Gamma_1 \cap \Gamma_2 = \emptyset$, in such a way that there is no interaction $J_{ij}$ connecting spins in the two
subsets, i.e. $J_{ij}=0$ if $i \in \Gamma_1$ and $j\in \Gamma_2$. 
The log partition function satisfies an additive property: if $\Gamma$ is disconnected, then 
$W_{\Gamma}(\mathcal{J}_{\Gamma})= W_{\Gamma_1}(\mathcal{J}_{\Gamma_1}) + W_{\Gamma_2}(\mathcal{J}_{\Gamma_2})$.
Thanks to this property, it is easy to show that $\Delta W_{\Gamma}=0$ for disconnected clusters.

Second, we note that $\Delta W_{\Gamma}$ becomes small when the size of the cluster grows. In fact, when the size of the cluster
becomes larger than the correlation length of the model, the far away spins have very weak interactions and the cluster behaves almost like
a disconnected cluster. In this situation, it is reasonable to expect that $\Delta W_{\Gamma}$ vanishes quickly.
Hence, we can truncate the series by selecting a list $L$ of small enough connected clusters, thus 
obtaining an approximation of the function $W_{\Omega}$:
\begin{equation} \label{eq:approx}
W_{\Omega} (\mathcal{J}) \sim \sum_{\Gamma \in L} \Delta W_\Gamma(\mathcal{J}_{\Gamma}) \ .
\end{equation}
The quality of the approximation of Eq.(\ref{eq:approx}) depends on the value of the neglected clusters, i.e. on the procedure used to construct the list $L$,
which we discuss in the next section.

\subsection{Adaptive Cluster Expansion: general construction and selection rules} 

The main problem of a cluster expansion is how to choose the list $L$ of clusters. One possibility is to sum them by size, i.e.
to consider all clusters whose length is lower than a given $K_{max}$. 
Because the calculation of the partition function grows exponentially with the size of the clusters,
 $K_{max}$ is fixed by computational limitations.
 Here, using a normal personal computer, we can reach a $K_{max}=18$ .
However,  clusters having the same length can have very different $\Delta W_{\Gamma}$ and clusters having different length can have comparable and sometimes
opposite in sign contributions, so 
this is not a well-converging procedure.

A more efficient one  
is to discard the clusters whose $\vert \Delta W_{\Gamma} \vert$ is smaller than a fixed threshold $\Theta$ (we will call them \textit{non significant} 
clusters)~\cite{Monasson-Cocco,Monasson-Cocco2}.
Clearly, even for rather small systems, it is not possible to compute all of the $2^N -1$ clusters $\Delta W_{\Gamma}$ to find the significant ones. Thus we use a recursive construction rule for the clusters (see the pseudocode of Algorithm \ref{alg:ACE}, readapted from Ref.~\cite{Monasson-Cocco2}).
We begin with the computation of $\Delta W_{(i)} = W_{(i)}= \log(\sum_{\sigma_i=0,1}  e^{\beta h_i \sigma_i})$ for all $N$ clusters of size $k = 1$.
Then we compute $\Delta W_{(i,j)}$ for all ${N \choose 2}$  clusters of size $k = 2$. From Eq.~\eqref{clust_exp}, that gives
$\Delta W_{(i,j)} = W_{(i,j)} - W_{(i)} - W_{(j)}$.
Each subsequent step follows the same pattern. Firstly, clusters with $\mid \Delta W_{\Gamma} \mid < \Theta $ are discarded and we add them to the list of \textit{non significant clusters}. 
We include in the next step of the expansion all clusters which are unions of two \textit{significant} clusters of size $k$, $\Gamma= \Gamma_1 \cup \Gamma_2$, such that the new cluster $\Gamma$ contains $k + 1$ spins. 
Note that this rule prevents any combinatorial explosion of the computational time, as it only selects a relatively small subset of the total number of clusters.
At the end of the procedure, 
we have a list $L$ of \textit{significant} clusters and we use them in the expansion of Eq.~(\ref{eq:approx}) to compute an approximation 
of the true free energy\footnote{Note that we considered also stricter versions of this rule, where a higher number of significant subclusters is required for a cluster to be significant, but we found
that such stricter choices prevent the inclusion of relevant clusters, so in the following we stick to this rule.}.

\label{sec:ACE}

\begin{algorithm}[t]
\caption{Adaptive Cluster Expansion}
\label{alg:ACE}
\begin{algorithmic}
\Require $N$, $\Theta$, routine to calculate $\Delta W_{\Gamma}$ from $\mathcal{J}_{\Gamma}$
\State $List \gets \emptyset$  All selected clusters 
\State $ListNonSignificant \gets \emptyset$ (The list of \textit{Non significant clusters})
\State $SIZE \gets 1$
\State $List(1) \gets$ add all clusters of $SIZE=1$
\While{$List(SIZE) != \emptyset$}
\State $List(SIZE+1) \gets \emptyset$
\For {each cluster $\Gamma \in List(SIZE)$}
		\State $Spinset \gets$ $\{$spins contained in $\Gamma \}$
	\EndFor
	\For{every spin $s \in SpinSet$ and every $\Gamma \in List(SIZE)$}
		\State $\Gamma' \gets \Gamma \cup s$
			\If {$\Gamma'$ contains $SIZE+1$ spins}
				\State add $ \Gamma' $ to $ListPossibleCluster$  
			\EndIf
	\EndFor

	\For{every $\Gamma' \in ListPossibleCluster$}
	\State \textbf{Construction  RULE} :
	\If {$\Gamma'$ appears at least twice in $ListPossibleCluster$ \textbf{and} $|\Delta W_{\Gamma'}| \geq \Theta	$}
		\State add $\Gamma'$ to $List(SIZE+1)$ 
	\ElsIf{$\Gamma'$ appears $\geq$twice in $ListPossibleCluster$ \textbf{and} $| \Delta W_{\Gamma'} | < \Theta$}
		\State		 add $\Gamma'$ to $ListNonSignificant$
	\EndIf

\EndFor
\State $SIZE \gets SIZE+1$
\EndWhile

\For {$\Gamma \in List$}
\State $W \gets W +\Delta W_{\Gamma}$
\State $\langle \sigma_i \rangle \gets \langle \sigma_i \rangle +\frac1\beta \frac{d }{dh_i} \Delta W_{\Gamma}$
\State $\langle \sigma_i \sigma_j \rangle \gets \langle \sigma_i \sigma_j \rangle + \frac1\beta\frac{d }{dJ_{ij}} \Delta W_{\Gamma}$
\EndFor
\Output $W,\langle \sigma_i \rangle, \langle \sigma_i \sigma_j \rangle, List$ of clusters
\end{algorithmic}
\end{algorithm}

Algorithm~\ref{alg:ACE}, which contains the main iteration loop that selects the clusters, calls as a subroutine Algorithm~\ref{Alg:deltaW}, which calculates the contribution $\Delta W_{\Gamma}$ of the cluster $\Gamma$ and, in turn, calls a subroutine, Algorithm~\ref{alg:free_entr}, to calculate the free energy of the clusters, $W_{\Gamma}$.
Note that the derivatives of the cluster free energies are required to compute the correlations according to Eq.~\eqref{eq:Corr} and Algorithm~\ref{alg:ACE}.
Explicit expressions of these derivatives can be obtained analytically, and their numerical evaluation requires only a minor modification of Algorithm~\ref{alg:free_entr}.
The time to calculate $W_{\Gamma}$, for a cluster of size $k$, is $O(2^{k})$, but the time to select clusters and compute $\Delta W_{\Gamma}$ is approximately linear in the number of clusters.

\begin{algorithm}[t]
\caption{Computation of $\Delta W_{\Gamma}$} \label{Alg:deltaW}
\begin{algorithmic}
\Require $\Gamma$(of size $k$), routine to calculate $W_{\Gamma}= \log Z_{\Gamma}$

$\Delta W_{\Gamma} \gets  W_{\Gamma}$
\For{$SIZE=k-1$  \textbf{to} $1$} 
	\For{Every $\Gamma'$ with $SIZE$ spins in $\Gamma$}
	$\Delta W_{\Gamma} \gets \Delta W_{\Gamma} - \Delta W_{	\Gamma'}$
	\EndFor
\EndFor
\Output{$\Delta W_{\Gamma}$}
\end{algorithmic}
\end{algorithm}

\begin{algorithm}[t]
\caption{Routine to calculate $W=\log Z$} \label{alg:free_entr}
\begin{algorithmic}

\Require cluster $\Gamma$, routine to get $\mathcal{J}_{\Gamma} = \lbrace h_i, J_{ij} \rbrace_{i,j \in {\Gamma}}$ , $SIZE$ of $\Gamma$

\State ${SPINS}\gets [\overbrace{0,0,0 \ldots, 0}^{SIZE}]$
\State $Z \gets 0$

\For {$i=1,2,\ldots, 2^{SIZE}$}
	\State $TEMP \gets 1$
	\For{$j=1,2,\ldots, SIZE$}
		\If{$SPINS[j]==1$}
		$TEMP \gets TEMP \cdot \exp[h_j]$
		\For	{$k=j+1,\ldots, SIZE$}
			\If{$SPINS[k]==1$}
				$TEMP \gets TEMP \cdot \exp[J_{jk}]$
			\EndIf
		\EndFor
		\EndIf
	\EndFor

\State $Z \gets Z+TEMP$

 \For{$j=1,2,\ldots, SIZE$} 			(function to flip the SPINS)
			\If{$SPINS[j]==1$} $SPINS[j]=0$
			\ElsIf{$SPINS[j]==0$}  $SPINS[j]=1$ and \textit{break};
			\EndIf
  \EndFor	
\EndFor
\Output{$W_{\Gamma}=\log Z$}
\end{algorithmic}
\end{algorithm}

\subsection{Significance threshold} 
\label{sec:theta}

Choosing a good value of $\Theta$ is of crucial importance: if we choose a too small $\Theta$ we will likely consider lots of long clusters and the computational time will increase.
On the other hand if we choose a too high $\Theta$ we will consider too few cluster and the estimated value of $W_{\Omega}$ will be far from the true value.
Because we do not know a priori the optimal value of $\Theta$, we run the algorithm several times decreasing progressively the threshold:
\begin{itemize}
\item Start with a large value of $\Theta$.
\item Once all possible new clusters are constructed calculate $M = \max_{\Gamma \in \overline{L}} \vert \Delta W_{\Gamma} \vert$,
where $\overline{L}$ is the list of all the non-significant clusters that have been constructed and tested in the previous run.
\item Set a new value $\Theta = 0.99 M$ of the significance threshold.
\end{itemize}
In this way, at each step we are sure that new clusters will be considered significant, and new clusters will be constructed.
In this paper, in order to investigate the convergence properties of the algorithm,
we stop the program when $M=0 $, which indicates that all the connected components of the system have been considered and no more significant clusters can be constructed; or, we set a minimal value of $\Theta_{min}$ and we stop the program when $\Theta < \Theta_{min}$. Obviously, in practical applications, one can stop the code as soon as 
convergence of the relevant observables is reached.

For ordered systems, as we will see in  more detail  in the following, 
some clusters give a zero free energy contribution because of symmetries.
As an example, it can be shown (see section \ref{sec:examples-ord}) that in ordered lattices with no external field and in the
Ising spin representation, non zero contributions for clusters containing more than 2 spins come only from  closed loops.
The construction rule must then be adapted,  for such ordered cases, 
in order to avoid that the expansion stops because the list of clusters of a given size is empty.

\subsection{Observables} 
 
In the following we report the results of our algorithm when applied to Ising models with different interaction structures.
Interesting quantities are:
\begin{itemize}
\item the approximate value of $W_{\Omega} =\log Z_{\Omega} (\mathcal{J})$ as a function of $\Theta$, given by $W_{\Omega}(\Theta)=\sum_{\Gamma \in L(\Theta)} \Delta W_\Gamma$, where the list $L(\Theta)$ includes all the significant clusters,
\item $N_{con}(k)$, the number of clusters of size $k$ constructed by the algorithm, and the total $N_{con} = \sum_k N_{con}(k)$,
 \item $N_{sel}$, the number of significant clusters, i.e. the clusters $\Gamma \in L(\Theta)$,
\item $K_{max}$, the size  of the largest significant clusters $\Gamma \in L(\Theta)$,
\item $N_{op}  = \sum_k N_{con}(k) 2^k$ which is an estimate of the number of operations performed by the algorithm, because constructing a cluster of size $k$
requires $2^k$ operations.
\end{itemize} 
We will examine the behavior of these quantities in
several examples of Ising models. Along the way, we illustrate several important 
properties of the cluster expansion.
Because most of the examples we consider are diluted, it is convenient to introduce the following notation.
The {\it interaction graph} ${\cal G}$ is such that a link $(ij) \in {\cal G}$ if and only if 
$J_{ij} \neq 0$. The size $\ell = | {\cal G} |$ is the number of non-zero couplings. Also, we call
$\partial i$ the set of neighbors of $i$, i.e. $\partial i = \{ j : J_{ij} \neq 0 \}$.

\section{Applications to ordered systems} 
\label{sec:examples-ord}

\subsection{One-dimensional Ising model without external field}

\begin{figure}[t]
\centering
\includegraphics[width=.8\textwidth]{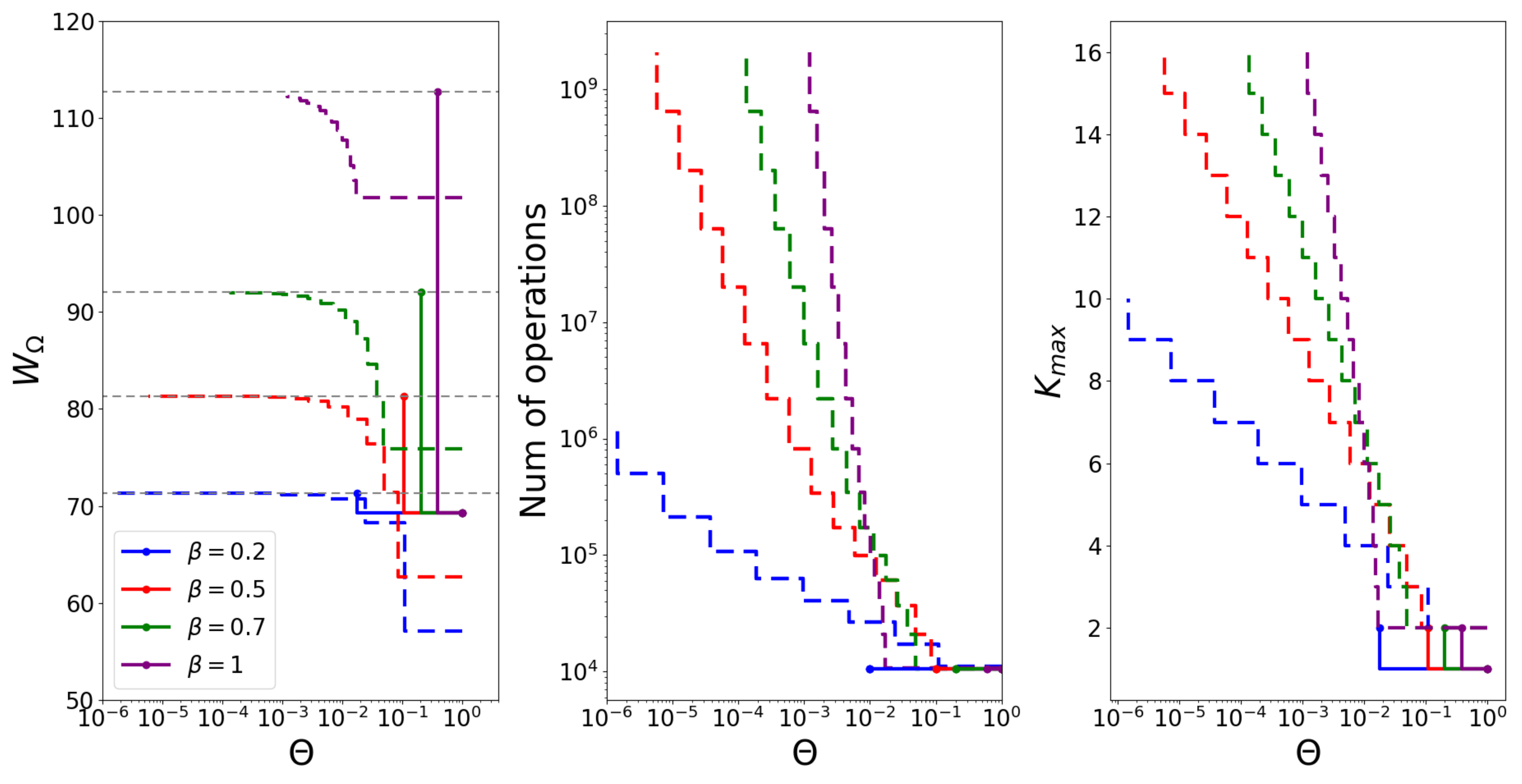}
\includegraphics[width=.8\textwidth]{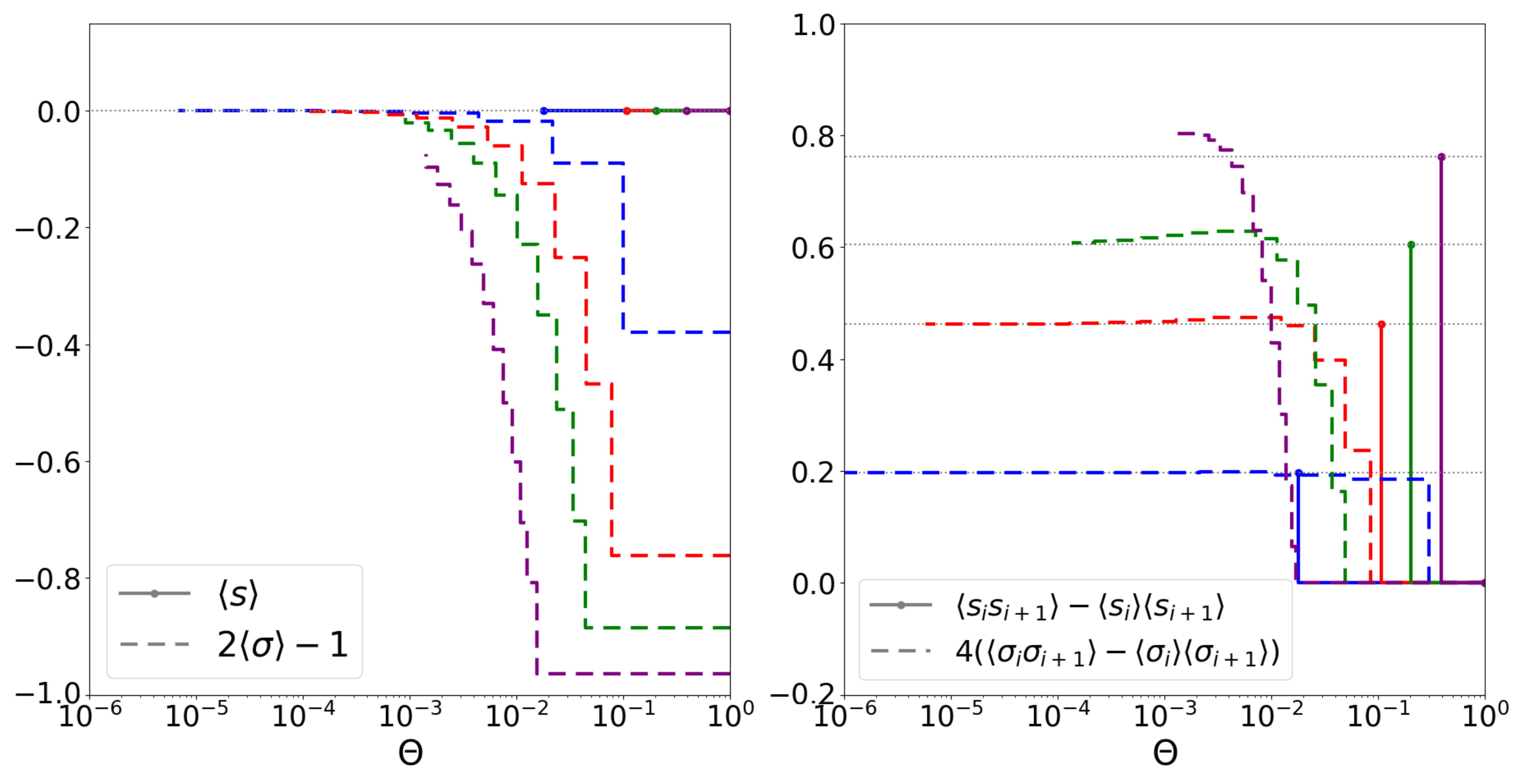}
\caption{Performance of the ACE for the one dimensional Ising chain with inversion symmetry,
at several values of temperature.
From top left to bottom right, the free energy, maximum cluster size, number of operations, 
magnetization, and nearest-neighbor correlations are plotted as function of the threshold $\Theta$.
The full lines are the results of the algorithm with Ising spins $s_i=\pm 1$, 
the dashed line whit Boolean spins $\sigma_i=0,1$.
The black  dotted line is the exact solution given by Eq.~\eqref{eq:1dWexact}. 
As discussed in the text, the Ising spin expansion converges using only clusters of length 1 and 2;
the Boolean spin expansion is much slower in this case.} 
\label{Ising1dim}
\end{figure}

\begin{figure}[t]
\centering
\includegraphics[width=.9\textwidth]{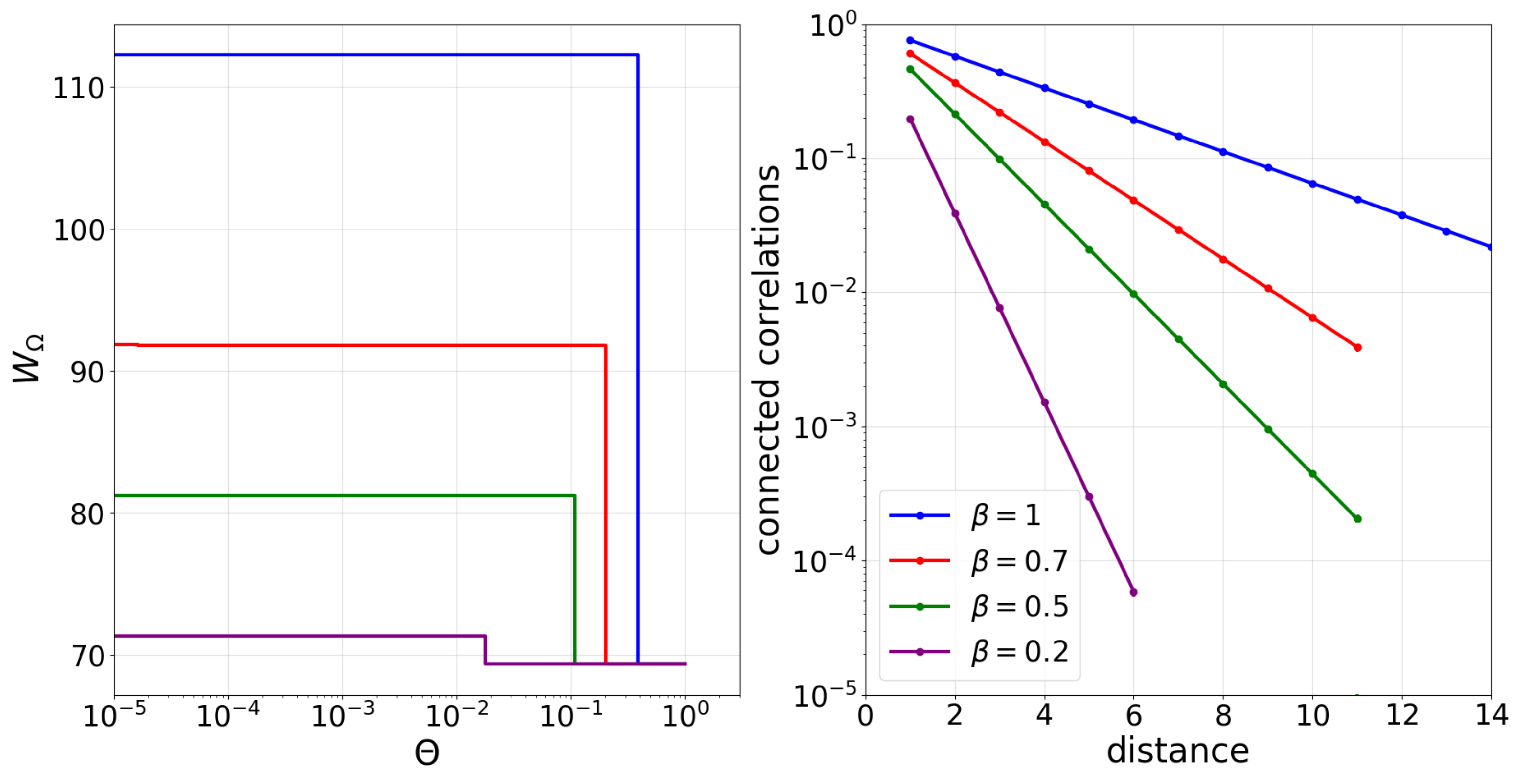}
\caption{Performance of the ACE for the one dimensional Ising chain with Ising spins $s_i=\pm 1$,
	when adding a small magnetic field $h=10^{-3}$. 
	This term does not alter the free energy with respect to the case without external field, shown in Figure~\ref{Ising1dim}.
	But, because 
	all clusters with more than two spins acquire an exponentially small 
$\Delta W$, it allows the algorithm to construct clusters of arbitrary length and thus to compute the $k$-spin connected correlation for $k>2$.}
	\label{Ising1dim_h_small}
\end{figure}

We first consider the one-dimensional Ising model with Ising spins and nearest-neighbour interactions, 
$J_{i,i+1} =J$, and zero external field, $h_i=0$. The spins are on a ring with $N$ sites (periodic boundary conditions). The Hamiltonian is:
\begin{equation}
H=- {J} \sum_{i=1}^N s_i s_{i+1} \ , \qquad s_{N+1}=s_1 \ , \qquad s_i \in \lbrace -1,1 \rbrace \ .
\end{equation}
Using Boolean variables $\sigma_i = (s_i + 1)/2$, the Hamiltonian becomes:
\begin{equation}
H=- {4 J} \sum_{i=1}^N \sigma_i \sigma_{i+1} + 4J \sum_{i=1}^N \sigma_i -NJ  \ , \qquad \sigma_{N+1}=\sigma_1  \ , \qquad \sigma_i =\lbrace 0,1 \rbrace \ .
\end{equation}
In Figure \ref{Ising1dim} we report results for the case $N=100$ and $J=1$ for different $\beta$.

The first observation is that, although the final result for $W_\Omega$ must be independent of the representation of the binary variables 
(Ising or Boolean), the intermediate steps of the cluster expansion strongly depend on the representation. The reason is that interactions
are classified differently in the two representations. In the Ising spin representation, the one-spin clusters have zero field and coupling,
hence $W_{(i)} = \log 2$. Also, in the Ising spin representation, all clusters keep the symmetry under inversion $s_i \to -s_i$ of the original Hamiltonian,
so the contribution to the average magnetization $\langle s_i \rangle$ is zero for all clusters.
On the contrary, in the Boolean representation the one-spin clusters have a field $h_i = - 4 J$, hence 
$W_{(i)} = \log \sum_{\sigma=0,1} e^{-4 \beta J} = \log (1+e^{-4 \beta J})$. Also, the one-spin clusters contribute to the
average a quantity $\langle \sigma_i \rangle = e^{-4 \beta J}/(1+e^{-4 \beta J}) < 0.5$. Despite the fact that the Hamiltonian
is globally invariant under $\sigma_i \to 1-\sigma_i$, the symmetry results from a cancellation between fields $h_i$ and couplings $J_{ij}$,
and is broken by the cluster expansion. While the cluster expansion in $s_i$ is an expansion around independent paramagnetic spins,
for large enough $\beta J$, the cluster expansion in $\sigma_i$ is an expansion around independent 
strongly magnetized Boolean spins with $\langle \sigma_i \rangle \sim 0$. For this reasons, in a system with inversion symmetry 
(which is not spontaneously broken), the Ising spin representation might seem more appropriate.

There is, however, another important consequence of the symmetry in the Ising-spin variable. 
In the one-dimensional case, 
clusters made by non-consecutive spin have $\Delta W=0$, while due to translational invariance the $n$ spins clusters are all identical.
The $\Delta W_{n}$ of a cluster made by $n$ consecutive spins is
$\Delta W_{n}= W_{n} - 2 \Delta W_{n-1} -3 \Delta W_{n-2} - \ldots$,
which implies
$\Delta W_{n} = W_{n} -2 W_{n-1} +W_{n-2}$ for $n>2$.
We have $Z_{1}= 2$, $Z_{2}=2(e^{\beta J} + e^{-\beta J})$, 
$Z_{3}=2(e^{\beta J} + e^{-\beta J})^2 =2^{2-1} Z_{2}^2$ and in general
$Z_{n}= 2^{2-n} Z_{2}^{n-1}$ for $n>2$.
Then
\begin{equation}
\Delta W_{n} =\log Z_{n} - 2 \log Z_{n-1} + \log Z_{n-2} =0 \ , \qquad \mbox{for } n>2 \ ,
\end{equation}
which shows that all clusters with $n\geq 3$ have vanishing $\Delta W$, except for the cluster made by all $N$ spins
 which has an additional contribution, exponentially small in $N$, coming from the $N$ consecutive spins. 
Summing  only one- and two-spin clusters is therefore accurate
with an exponentially small error in $N$. Moreover, as we will see in another example below, 
the fact that all three-spin clusters have $\Delta W=0$ can
become a problem, if there are higher order clusters that have non-zero $\Delta W$. 
Therefore,
we conclude that all clusters with $n>2$ spins have $\Delta W=0$, and the program will stop
after constructing the two-spin clusters.
Finally, it is important to stress that 
in order to obtain $k$-spin  connected correlations we need clusters of size $k$ even if the free energy has already reached convergence;
in fact, these clusters give zero contribution to the free energy, but a finite contribution to spin correlations. Our selection rule on the free
energy is then not appropriate to compute correlations in this case.

As shown in  Figure \ref{Ising1dim_h_small}, one possibility to solve this problem is to add a small magnetic field to the Hamiltonian.
In this way all clusters with $n>2$ acquire a $\Delta W$ which, while being exponentially small, is not zero anymore.
Therefore the program will not stop after constructing the two-spin clusters, but can reach clusters of arbitrary dimension $k$ and obtain the $k$-spin connected correlations.

Note that the above result is a simple instance of a more general result for binary variables on ordered interaction graphs.
Using the identity
\begin{equation}
e^{\beta J s_i s_j}= \cosh(\beta J) +s_i s_j \sinh(\beta J)= \cosh(\beta J) (1+w s_i s_j)  \ , \qquad w= \tanh(\beta J) \ ,
\end{equation}
the partition function can be rewritten in the following way (which is essentially a high-temperature expansion):
\begin{equation}
 Z=\sum_{\bf s} e^{\sum_{(ij)} \beta J s_i s_j} =  \cosh(\beta J)^{\ell} \sum_{\bf s} \prod_{(ij)} (1+w s_i s_j) \ ,
\end{equation}
where $\ell$ is the total number of links $(ij)$ carrying a non-zero coupling $J$.
Furthermore, the sum of a product of spins is
\begin{equation}
 \sum_{\bf s} s_{i_1} s_{i_2} \ldots s_{i_n} = 
\begin{cases}
2^N  & \mbox{if all spins appear an even number of times} \ , \\
0  & \mbox{otherwise} \ .
\end{cases}
\end{equation}
Developing the product $\prod_{(ij)} (1+w s_i s_j)$ one obtains products of spins over links. The only way to have each spin appearing twice is
to form a closed polygon. Hence
\begin{equation}
\begin{split}\label{eq:Zpolygon}
Z&=2^N \cosh(\beta J)^\ell \left[ 1+\sum_{p>0} w^{p} g_N(p)\right] \ , \\
W_\Omega&=\log Z = N \log 2 + \ell \log\cosh(\beta J) + \log \left[ 1+\sum_{p>0} w^{p} g_N(p)\right] \ ,
\end{split}\end{equation}
where $g_N(p)$ the number of closed polygons in $\cal G$ having perimeter $p$.
Specializing this to a one-dimensional ring of $N$ sites, we have $N$ links and the only closed polygon is the ring itself, with perimeter $p=N$, 
so that
\begin{equation}\label{eq:1dWexact}
W_\Omega= N \log 2 + N \log\cosh(\beta J) + \log(1+ w^N) \ .
\end{equation}
The $N$ one-spin clusters contribute $W_1 = \log 2$, which is the first term. The $N$ two-spin clusters made of neighboring spins contribute
$\Delta W_2 = \log \sum_{s_1 s_2} e^{\beta J s_1 s_2} - 2 W_1 = \log[2 (e^{\beta J} + e^{-\beta J})] - 2 \log 2=  \log \cosh(\beta J)$, 
which is the second term. 
The ring contributes the last term, $W_N = \log(1+ w^N)$, which vanishes exponentially for $N\to\infty$.

In summary, as shown in Figure~\ref{Ising1dim}, for a one dimensional system with inversion symmetry:
\begin{itemize}
\item The Ising spin representation converges extremely fast: only one-spin and two-spins clusters are computed, after which the expansion
stops. The error in the free energy is due to the $N$-spin clusters, which gives a contribution exponentially small in $N$. The magnetization
is zero at all orders, but only nearest-neighbor correlations can be computed.
\item The Boolean spin representation is slower, because it breaks the inversion symmetry, and it is an expansion around a magnetized state.
One has to include several clusters for the free energy and the magnetizations to converge to their asymptotics value.
However, in this case one can compute all the $n$-point correlations, provided a sufficient number of clusters is included.
\end{itemize}

\subsection{One-dimensional Ising model with homogeneous external field}
\begin{figure}[t]
\centering
\includegraphics[width=.9\textwidth]{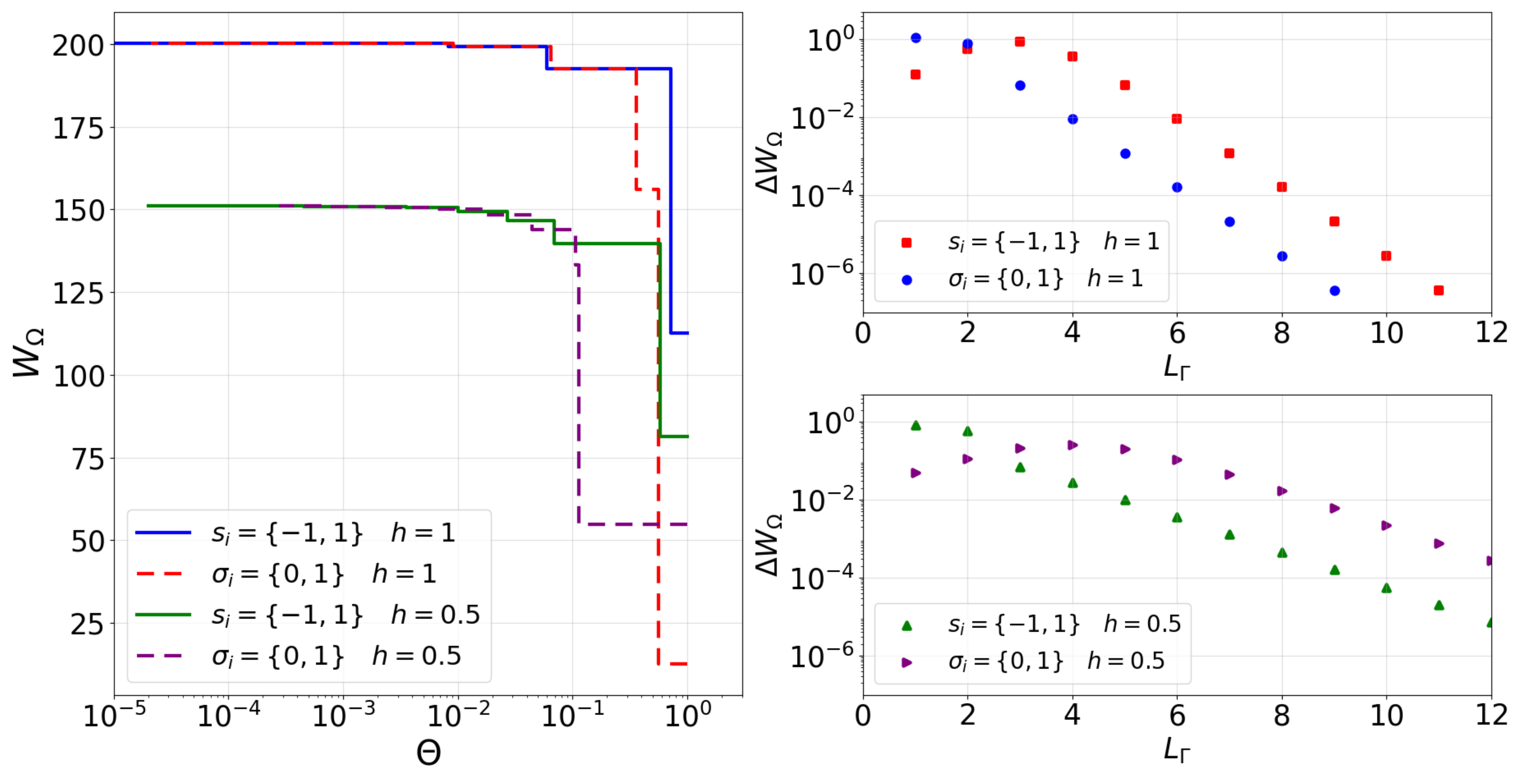}
\caption{Performance of the ACE for the one dimensional Ising model with
homogeneous field. On the left we plot the free energy, while on the right the
cluster contribution $\Delta W_{\Gamma} $ are plotted versus the length of the clusters. We
note that $\Delta W_{\Gamma} \sim \exp[-L_{\Gamma} /\xi ]$ where $\xi$ is the
correlation length of the system. The cluster contribution in the Boolean
representation does not decrease monotonically due to the external
field $2(h-2J)$. }
\label{Ising_1d_homogeneous}
\end{figure}

 The inversion symmetry (and the consequent cancellation of cluster contributions) of the previous model is broken if we add an external
field to the system.
 We then study the case of a homogeneous external
field, 
\begin{equation}
    H= - J \sum_{i=1}^N s_i s_{i+1} - h\sum_{i=1}^{N} s_i
\end{equation}
which in the Boolean representation becomes,
\begin{equation}
    H=- 4 J\sum_{i=1}^N \sigma_i\sigma_{i+1} -  2 ( h - 2J)  \sum_{i=1}^N \sigma_i + hN-NJ \ . 
\end{equation}
Results for this system are reported in Figure~\ref{Ising_1d_homogeneous} for $\beta=1$, $N=100$ and
two values of $h$. 
 The cluster contribution  $\Delta W_{\Gamma}$ decreases exponentially with the
 length of the clusters and the correlation length of the system:
$\Delta W_{\Gamma} \sim \exp[-L_{\Gamma} /\xi ] $ where $\xi \approx 0.49$ for
$h=1$ and $\xi \approx 0.96$ for $h=0.5$.

\subsection{Two-dimensional  Ising model}
\label{sec:Ising2d}

\begin{figure}[t]
\centering
\includegraphics[width=0.8\textwidth]{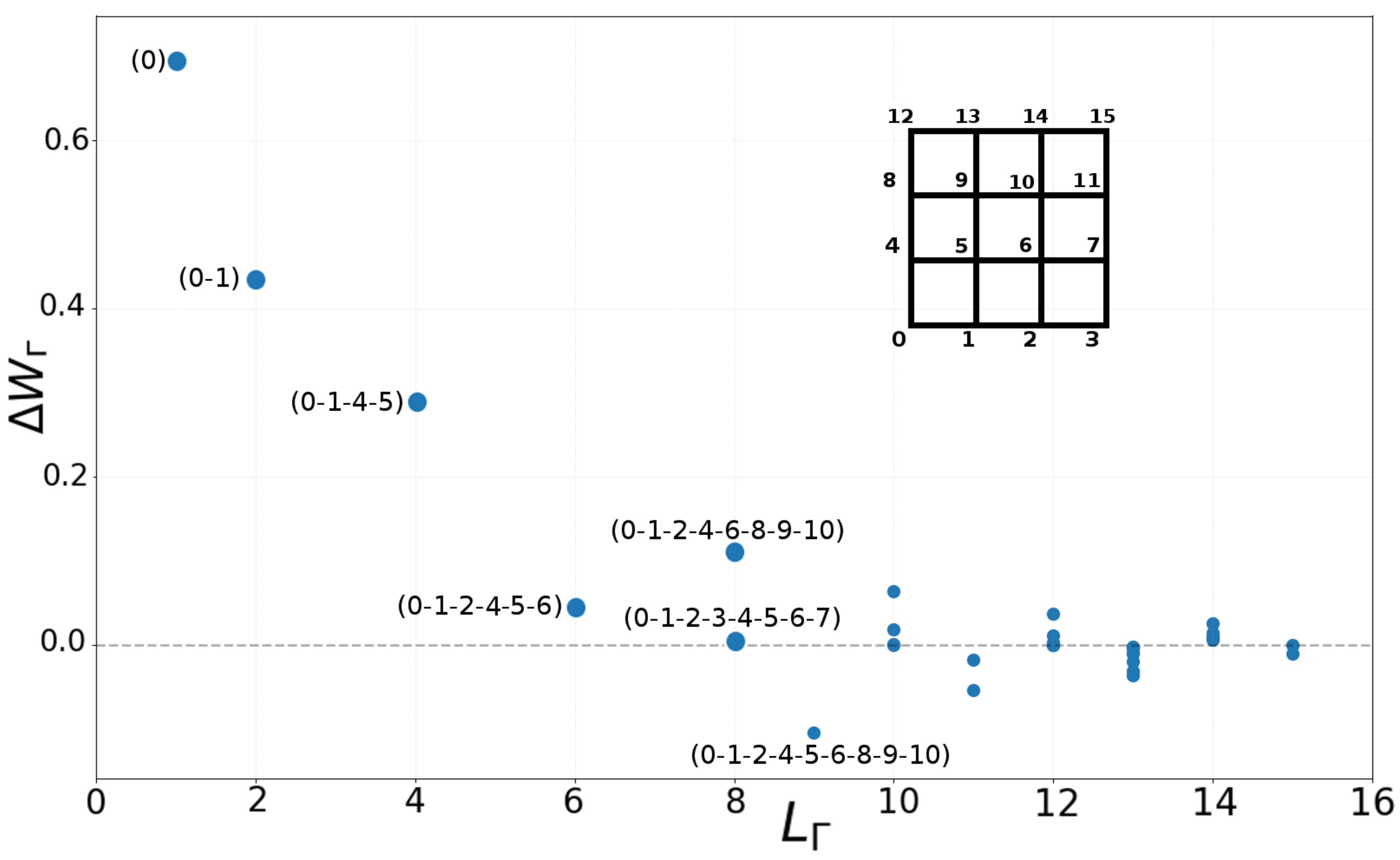}
\caption{ The $\Delta W_{\Gamma}$ which are different from zero, as a function of the size of the cluster, for a system of 4 $\times$ 4 spins on a square lattice with open 
boundary conditions. We note that $\Delta W_{\Gamma}$ is different from zero only for one- and two-clusters, or for closed polygons.
Furthermore, clusters with the same topology have the same $\Delta W_{\Gamma}$.}
\label{contributo_cluster_ising2d}
\end{figure}

\begin{figure}[t]
\centering
\includegraphics[width=0.8\textwidth]{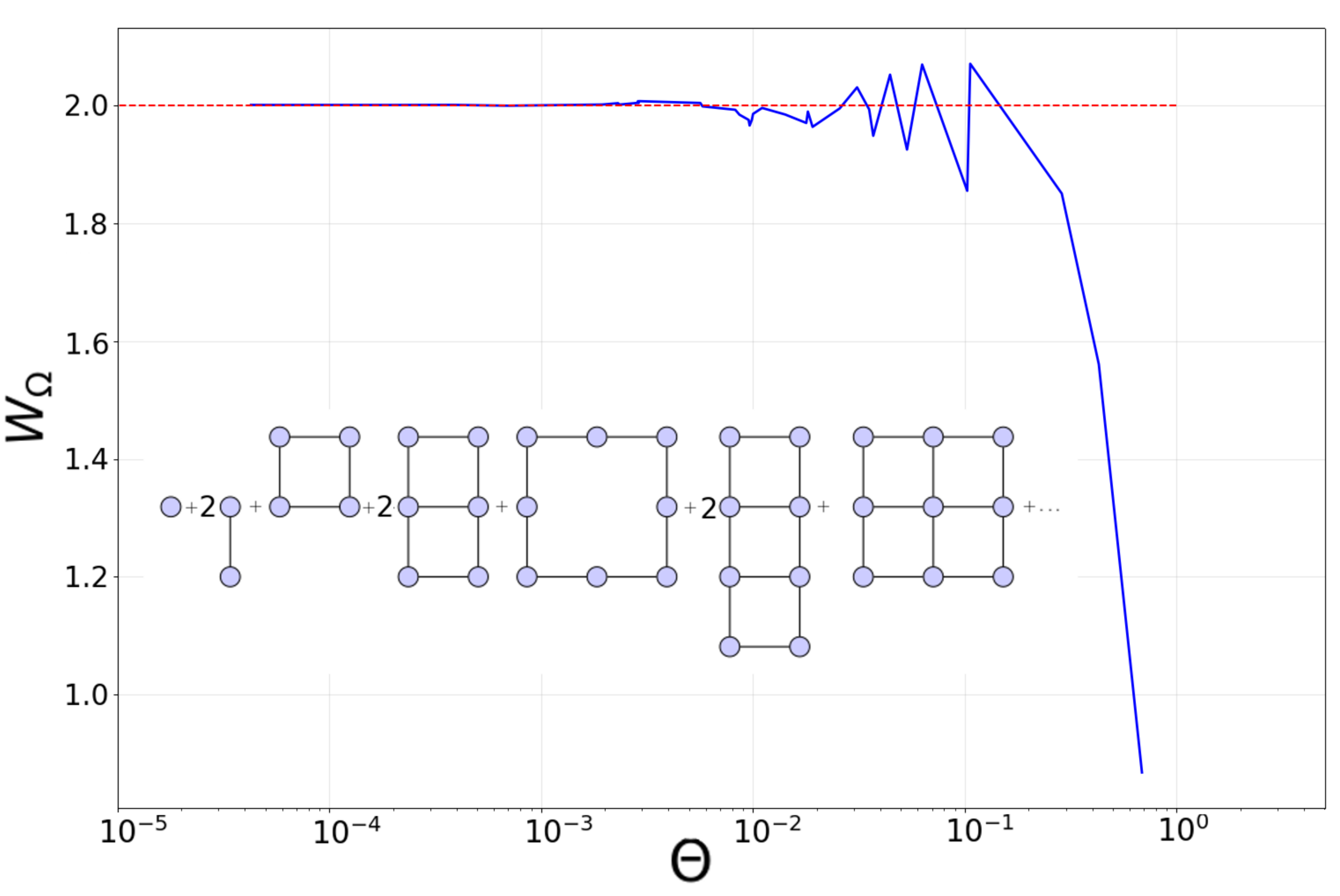}
\caption{ The list of clusters having a non zero contribution for a 2d Ising model on a square lattice, 
	at inverse temperature $\beta = 1$, together with their multiplicity $k_S$ 
(for the rotational symmetry). Their $\Delta W$ has been calculated using a $4 \times 4$ system.
The main panel gives an estimate of the free energy, obtained by summing up the significant clusters up to size 15, 
$W_\Omega(\Theta) = \sum_{\Gamma : |\Delta W_\Gamma|>\Theta }^{\vert \Gamma \vert \leq 15} k_S(\Gamma) \Delta W_{\Gamma}$, we get $W/N = 2.0031 \ldots$ close to the exact value in the thermodynamics limit $2.00035 \ldots$ (the red dotted line in the figure). }
\label{cluster_lista_is2d}
\end{figure}

We now study the ferromagnetic 2d Ising model on a square lattice with open boundary conditions.
The couplings are $J_{ij}=J$ if the distance between the spin $i$ and the spin $j$ is one and zero otherwise, and the fields are absent, $h_i=0$,
so the model has inversion symmetry.
The Hamiltonian is: 
\begin{equation} \label{H_ising2d}
 H=-{J} \sum_{(i j)} s_i s_{j} \ , \qquad s_i= \lbrace -1,+1 \rbrace \ .
\end{equation}
This model
has been solved by Onsager in the limit $ N \rightarrow \infty$~\cite{baxter}. 
The free energy and magnetization are given by:
\begin{equation}\label{eq:Ons}
\begin{split}
& k \doteq \frac{1}{\sinh\left(2 \beta J\right)^2} \\
& W=  -N \beta f = N \left( \frac{\log(2)}{2} + \frac{1}{2\pi}\int_{0}^{\pi}\log\left[\cosh\left(2 \beta J\right)^2+\frac{1}{k}\sqrt{1+k^{2}-2k\cos(2\theta)}\right]d\theta \right) \\
& m = \langle s_i \rangle= [ 1 - \sinh^{-4}(2\beta J) ]^{1/8} \ ,
\hskip30pt
\langle \sigma_i \rangle = (1+m)/2 \ .
\end{split}\end{equation}
In Figure~\ref{contributo_cluster_ising2d} we plot the cluster contributions $\Delta W_{\Gamma}$ as a function of the cluster size. 
Because the model has inversion symmetry, the polygon expansion in Eq.~\eqref{eq:Zpolygon} applies and indeed we observe
that only clusters formed by closed polygons have a non zero $\Delta W$.
The figure illustrates that our algorithm for Ising spins does not work well in this case.
Indeed, new clusters are created by adding a new spin to significant clusters (i.e. with $ \vert \Delta W \vert >  0$), 
but because all clusters with 3 spins have $\Delta W=0$ (so they are never significant) we cannot create larger clusters using our construction rule.
Consequently the expansion is interrupted after clusters of size one and two are constructed, which is not enough to obtain a good value for the free energy.
Clearly, this is a pathology due to the particular structure of this problem. The solution is to take into account the clusters having the largest contributions
$ \vert \Delta W \vert$, see Figure~\ref{contributo_cluster_ising2d}, with their multiplicity, depending on the system size and of the symmetries. In this way,
one can obtain very accurate estimations of the free energy even in the thermodynamic limit, as illustrated in Figure~\ref{cluster_lista_is2d}.

\begin{figure}[t]
\centering
\includegraphics[width=.9\textwidth]{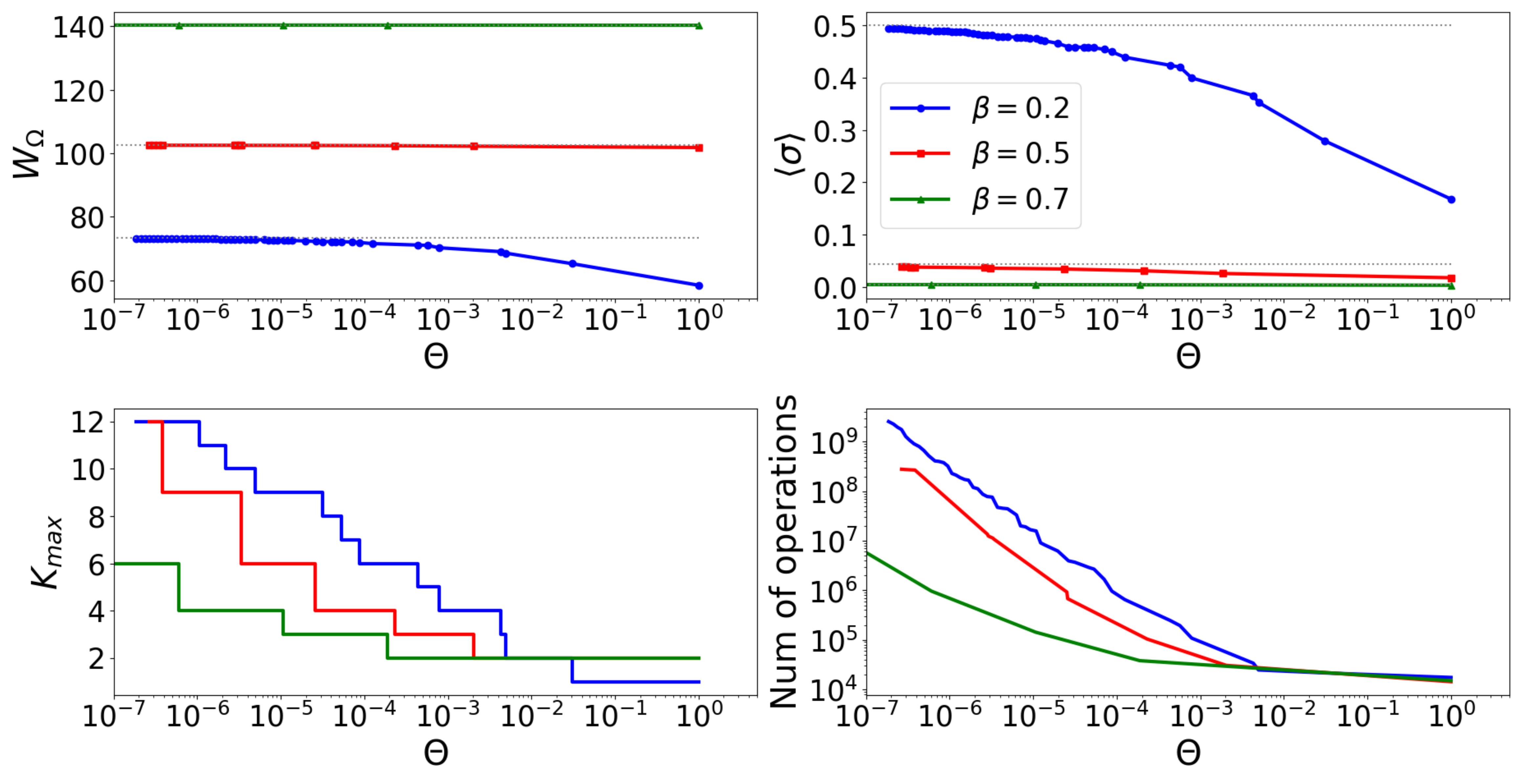}
\caption{Performance of the ACE on the two-dimensional ferromagnetic Ising model without external field, in the Boolean spin representation.
The results are compared with the exact Onsager solution (dashed lines). }
\label{Ising2dim}
\end{figure}

\begin{figure}[t]
\centering
\includegraphics[width=.9\textwidth]{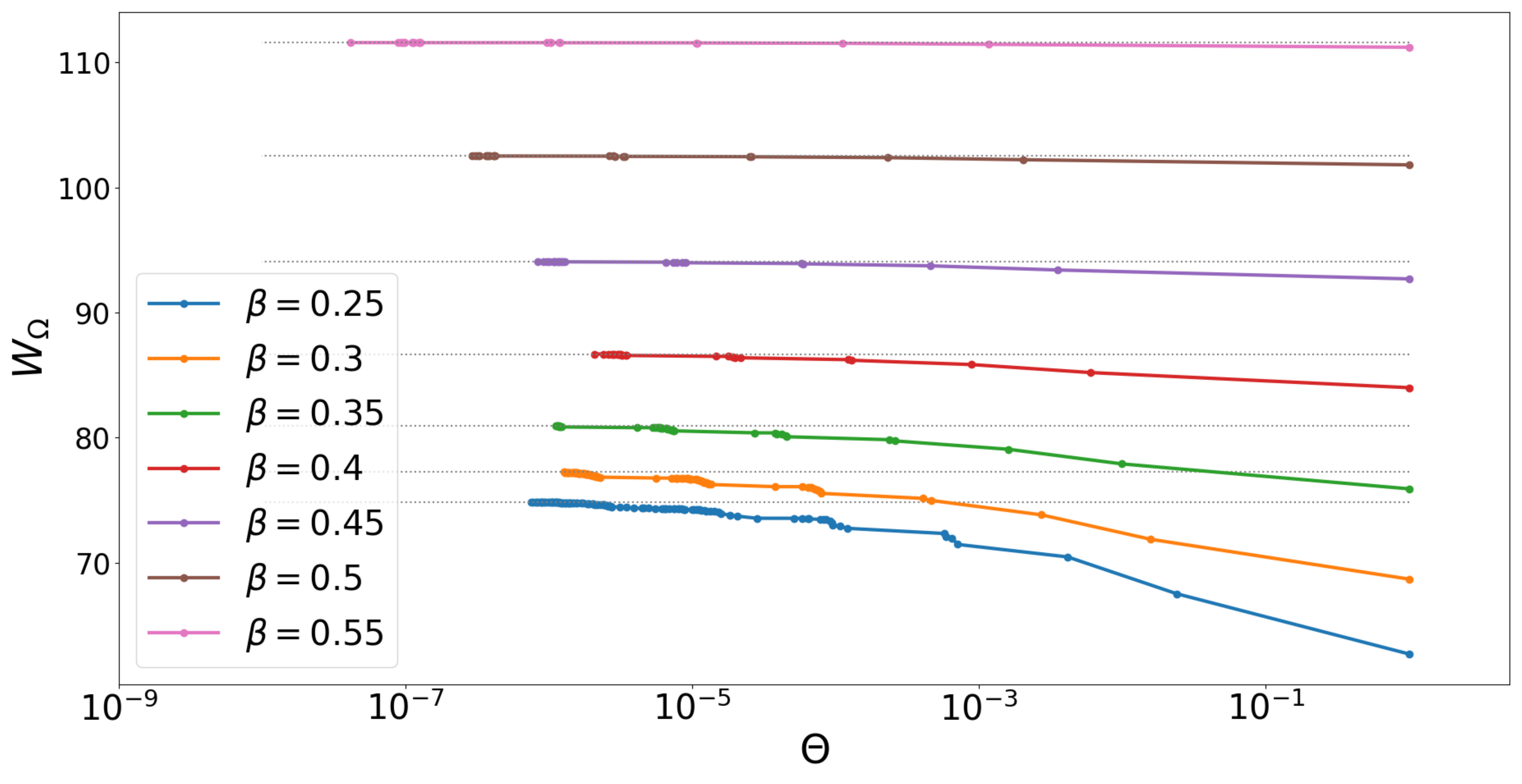}
\caption{Performance of the ACE on the two-dimensional ferromagnetic Ising model without external field, in the Boolean spin representation for temperatures close to the phase transition temperature $\beta_c \approx 0.44$. The dashed lines are the exact Onsager solution. }
\label{Ising2dim_critical}
\end{figure}

As in the one-dimensional case, another way to solve this problem can be to study the same system but using 
Boolean spins, for which the Hamiltonian in Eq.~\eqref{H_ising2d} becomes
\begin{equation}\label{eq:4}
 H= -4{J} \sum_{(ij)} \sigma_i \sigma_j +8J \sum_i \sigma_i -2NJ \ , \qquad \sigma_i= \lbrace 0,1 \rbrace \ .
\end{equation}
The inversion symmetry is then broken by the cluster expansion, and there are significant clusters
of any size.
In Figure~\ref{Ising2dim} we show the results for a system with $N=10 \times 10$ spins on a square lattice, with periodic boundary conditions, 
$J=1$ and $\beta=0.2,0.5,0.7$ (the Onsager critical point is at $\beta_c \sim 0.44$),
and we compare the free energy and the magnetization with the Onsager solution, Eq.~\eqref{eq:Ons}.
Note that the expansion is around the state with all $\sigma \sim 0$, 
because for the Hamiltonian in Eq.~(\ref{eq:4}), the one-spin clusters have a strong field $h=-8J$ that favors the $\sigma=0$ configuration.
Therefore, the expansion works better in the ferromagnetic phase, where the free energy and the magnetization converge immediately to one of the two possible values,
because the system is very strongly magnetized.
On the contrary, in the paramagnetic state where one has to include more clusters to reach zero magnetization.
As shown in Figure \ref{Ising2dim_critical}, the algorithm also properly converges when approaching the phase transition temperature $\beta_c \approx 0.44$.

\begin{figure}[t]
\centering
\includegraphics[width=.9\textwidth]{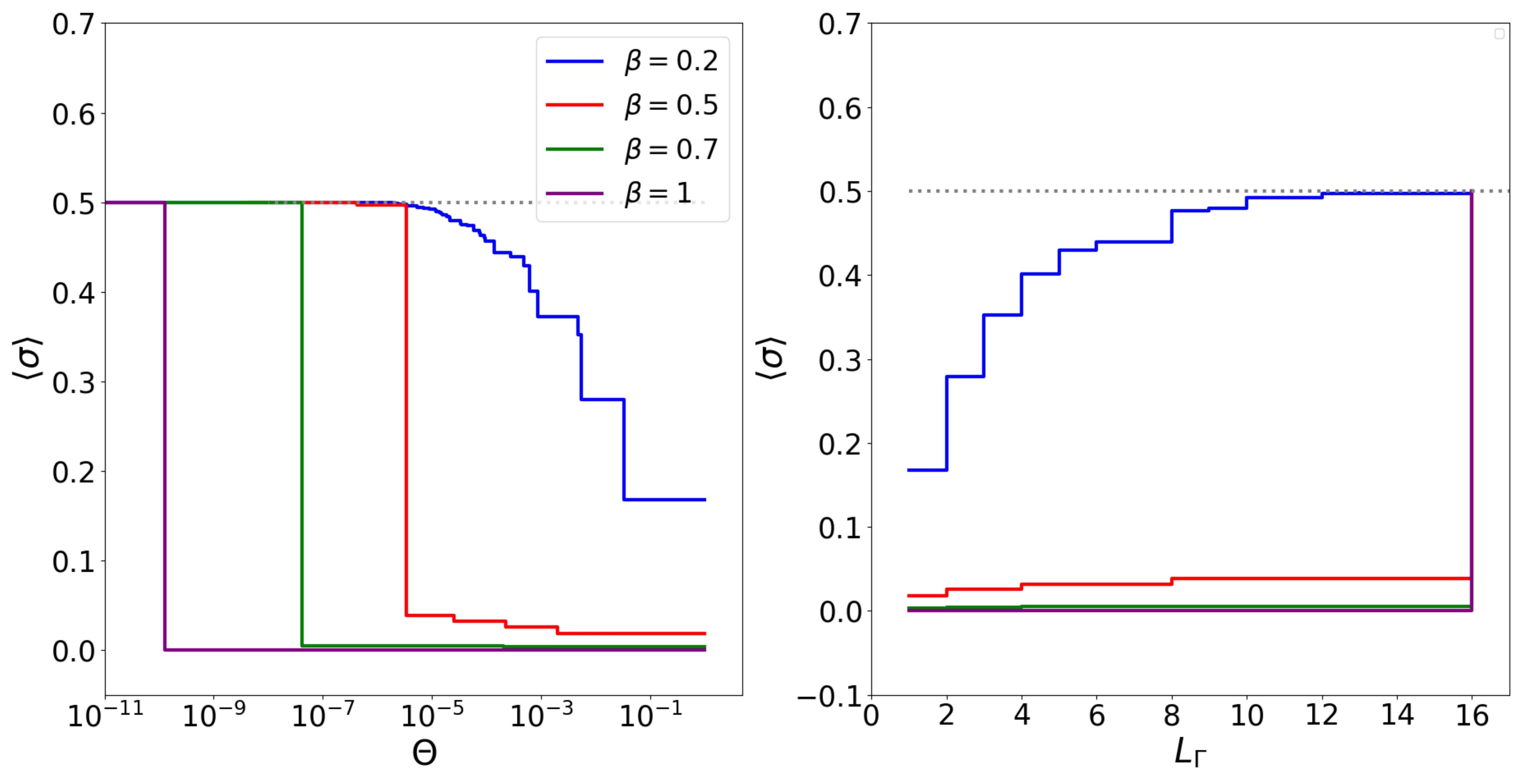}
\caption{Magnetization  $\langle \sigma \rangle$ as a function of $\Theta$ and of $L_{\Gamma}$ (the length of the cluster) for a $4 \times 4$ ferromagnetic
system without external field, in the Boolean representation as in Figure~\ref{Ising2dim}.
The value $\langle \sigma \rangle =0.5$ is recovered only by including the cluster with 16 spins.}
\label{magn_finite}
\end{figure}

Note that Osanger's solution is only correct in the thermodynamic limit ($N \rightarrow \infty$), where the inversion symmetry can be spontaneously broken, while
at finite $N$ one must always find $\langle \sigma \rangle=0.5$ for any $\beta$. However, zero magnetization is recovered only by considering very large cluster 
(of size comparable with the system size). To show this,
in Figure \ref{magn_finite} we plot the magnetization $\langle \sigma \rangle$ as a function of $\Theta$ for a small system made by $4 \times 4$ spins;
it is observed that the magnetization converges to $\langle \sigma \rangle < 0.5$, but when the cluster with 16 spins is included the symmetry is restored
and it jumps to $\langle \sigma \rangle = 0.5$.

\subsection{Summary}

The main message we learn from the application of the ACE method to ordered systems is that, when the correlation length is finite,
the cluster contribution 
${\Delta W_{\Gamma} \sim \exp[-L_{\Gamma} /\xi ]} $ decays exponentially with the cluster size.  Hence, the free energy converges
exponentially fast (in the size of the largest clusters) to the thermodynamic limit.
However, some symmetries can impose the vanishing of some $\Delta W_{\Gamma}$, which can spoil the efficiency of the construction
rule; symmetries should then be carefully taken into account.
Finally, both the speed of convergence and the role of symmetries depend on the spin representation (Ising or Boolean).

\section{Applications to disordered systems} \label{sec:examples-disord}

We now discuss the performances of the ACE when applied to disordered systems.
Note that while in many spin glass studies, external fields are absent and the inversion symmetry is thus preserved,
we prefer to consider the most general case in which external fields are present. In this way, the inversion symmetry is explicitly broken
and the problems discussed in section~\ref{sec:Ising2d} are not present. We also note that in applications to biological systems,
external fields are always present.

\subsection{One-dimensional model with random fields}

 \begin{figure}[t]
\centering
\includegraphics[width=.9\textwidth]{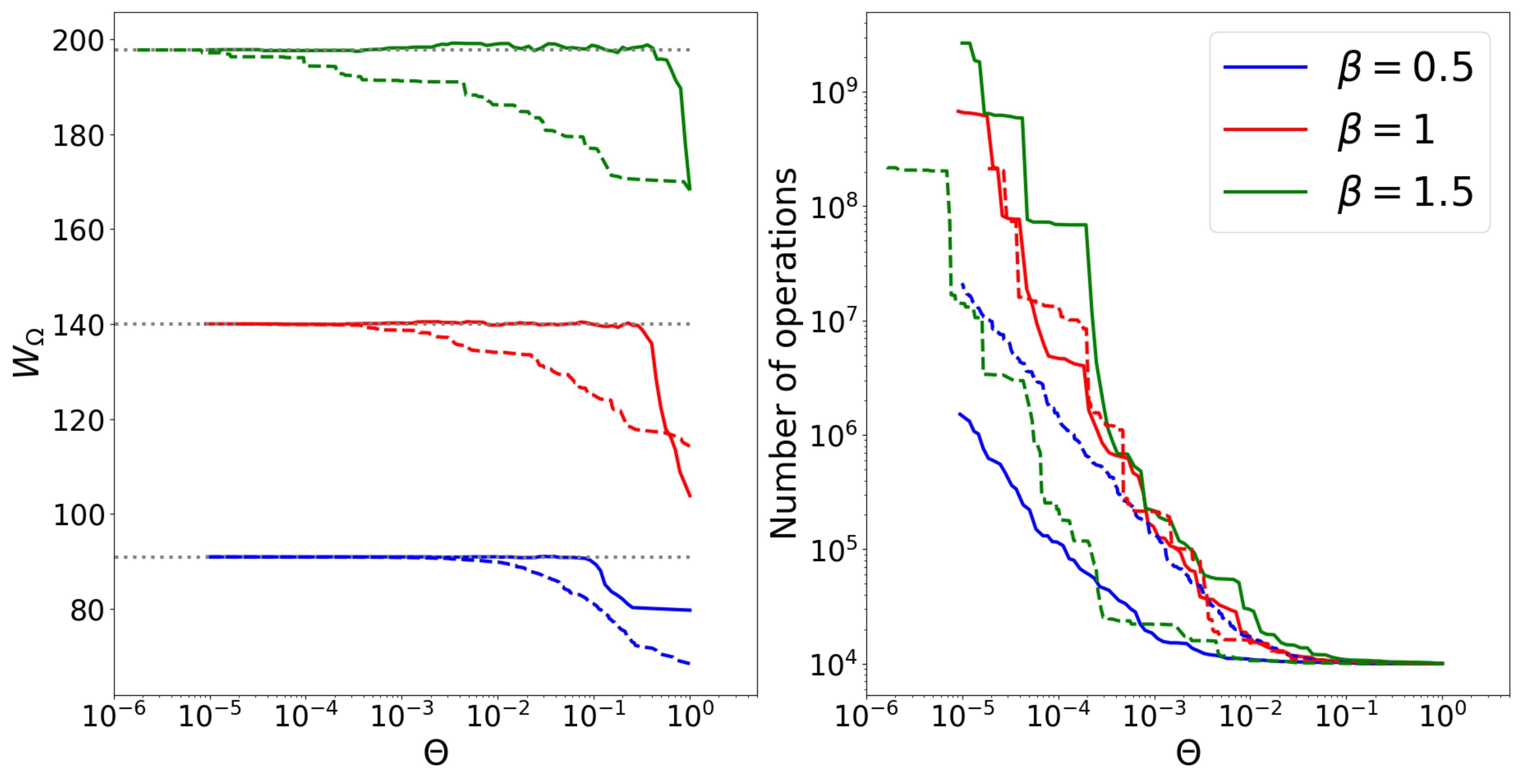}
\includegraphics[width=.9\textwidth]{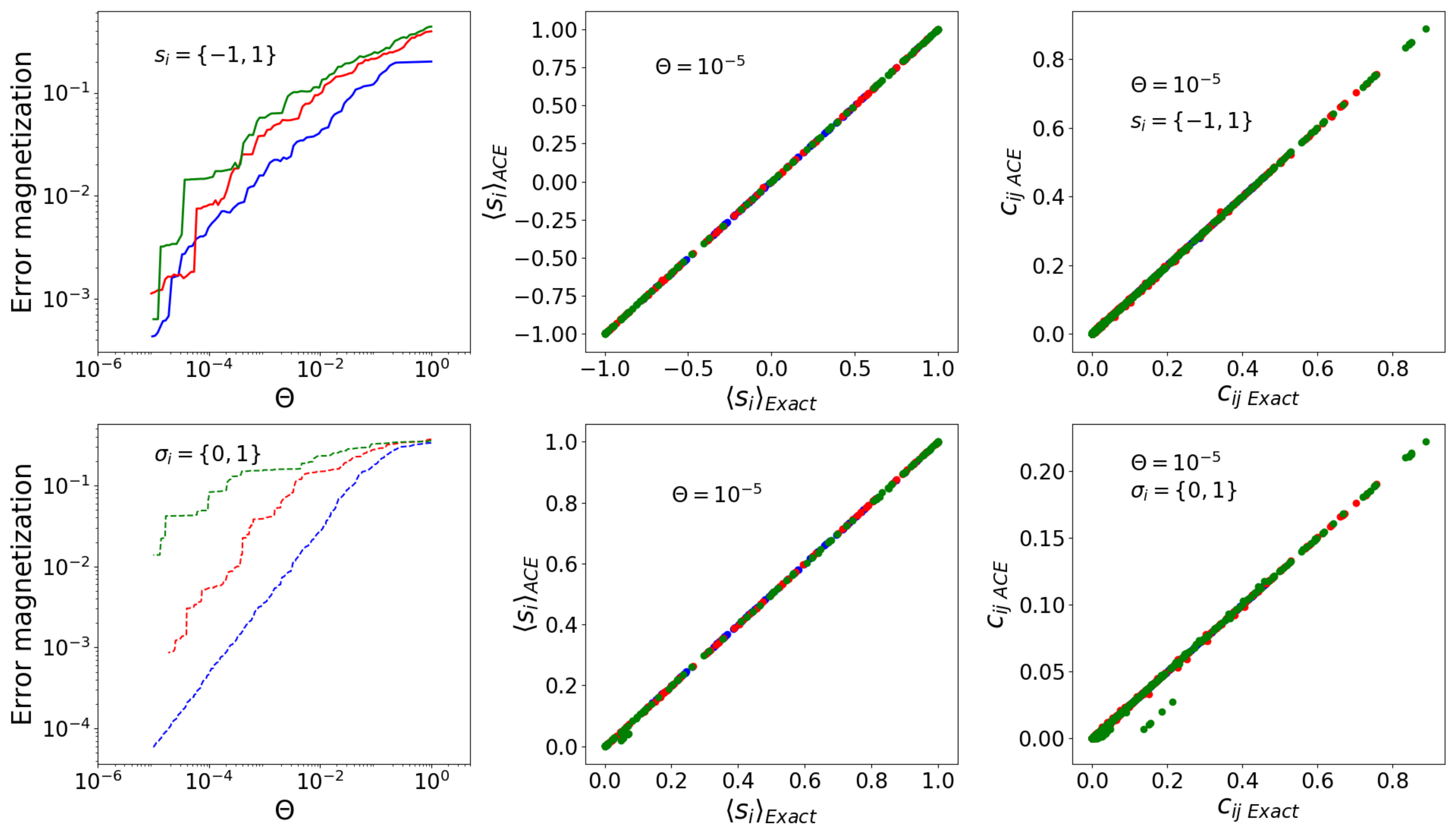}
\caption{
Performance of the ACE for the one dimensional Random Field Ising model,
at several values of temperature.
In the top row, the free energy and number of operations are plotted as function of the threshold $\Theta$.
The full lines are the results of the algorithm with Ising spins, the dashed line with Boolean spins. The black dashed lines are the exact results obtained by the transfer matrix method. 
In the bottom left plots, the mean error for the magnetization,
$ \frac{\sum_i |\langle \sigma_i \rangle (\Theta) - m_i|}{N}$ with $m_i = \langle \sigma_i \rangle$ being the exact magnetization, is plotted as a function
of $\Theta$. In the bottom right plots, the estimated magnetizations and connected correlations 
$c_{ij}= \langle \sigma_i \sigma_j \rangle - \langle \sigma_i \rangle  \langle \sigma_j \rangle$
 for $\Theta=10^{-5}$ are plotted versus the exact ones obtained by
the transfer matrix method.
}
\label{fig:1dRFIM}
\end{figure}

We now test the algorithm on a system with the same geometry, 
$N=100$ spins, ferromagnetic nearest-neighbour interactions, but random external fields. In the Ising spin representation, 
the Hamiltonian is:
\begin{equation} \label{RFM_hamiltonian}
H= - J\sum_{i=1}^N s_i s_{i+1} - \sum_{i}^{N} h_i s_i \ , \qquad  h_i \sim \mathcal{N}(0,1) \ ,
\end{equation}
where ${\cal N}(0,1)$ denotes a Gaussian random variable with zero mean and unit variance.
With Boolean spins, Eq.~(\ref{RFM_hamiltonian}) becomes:
\begin{equation}
H=\sum_{i=1}^N (4J-2h_i) \sigma_i - 4 J \sum_{i=1}^N \sigma_i \sigma_{i+1} + NJ - \sum_{i=1}^N h_i \ .
\end{equation}
The exact solution of this model can be obtained by the transfer matrix method.

In this case, both with Ising and Boolean spins, the algorithm constructs clusters with arbitrarily large number of spins
upon lowering $\Theta$ (Figure~\ref{fig:1dRFIM}). As in the case without random fields, we observe that the free energy
expansion converges faster with Ising spins than with Boolean spins. However, when looking at the magnetizations and correlations,
the convergence of the Ising expansion is only slightly better than the Boolean expansion. Both expansions achieve very good precision
in reproducing the magnetizations and correlations when $\Theta \sim 10^{-5}$.

 \begin{figure}[t]
\centering
\includegraphics[width=.9\textwidth]{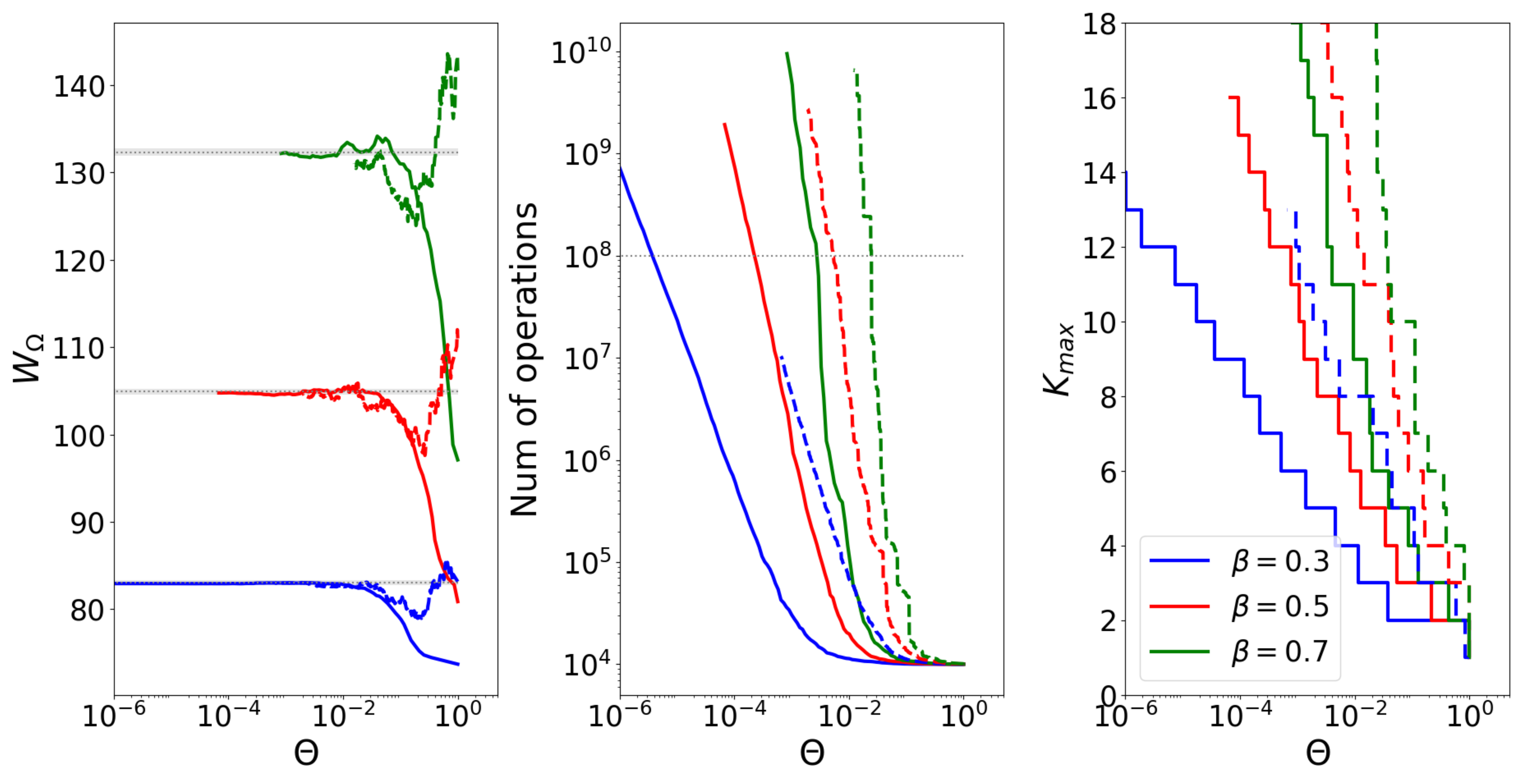}
\includegraphics[width=.9\textwidth]{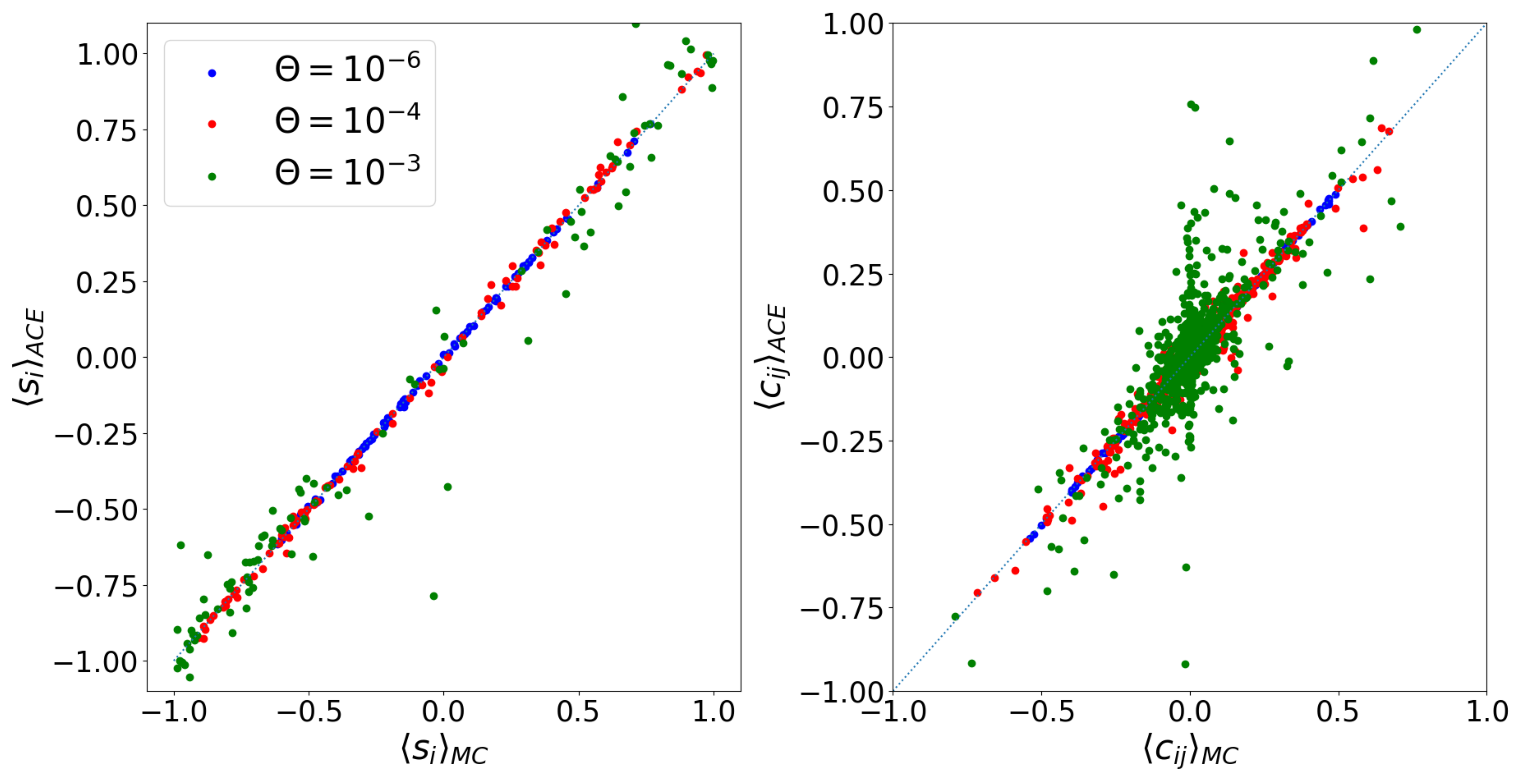}
\caption{Performance of the ACE for the two-dimensional Edwards-Anderson model with random fields. 
The full lines are the results using the Ising spin representation, while the dashed lines are the results using the Boolean spin representation.
On the upper left, the black dotted line is the free energy obtained using a Monte Carlo-Metropolis algorithm,
averaged over 10 repetitions of the simulation (the shaded error bar being the standard deviation), 
each requiring a computational time $\mathcal{O}(10^8)$ (the dashed line in the upper center plot).
In the bottom panels, the magnetizations $\langle s_i \rangle$ and connected correlations $c_{ij} = \langle s_i s_j \rangle -\langle s_i \rangle \langle s_j \rangle$ obtained from the ACE are compared with those obtained from Monte Carlo,
for the same three temperatures as in the upper panels, choosing at each temperature the smallest available $\Theta$.
}
\label{EAG}
\end{figure}

\subsection{Two-dimensional Edwards-Anderson spin glass model}

We also applied our algorithm 
to a disordered Ising model on a two-dimensional square lattice,
 with periodic boundary conditions, the so-called Edwards-Anderson spin-glass model with random external fields~\cite{fisher,mpv}. 
 The Hamiltonian is
\begin{equation} \label{Hamiltonian_EA}
H= - \sum_{(ij)} J_{ij} s_i s_j - \sum_{i=1}^{N} h_i s_i  \ , \qquad s_i=  \lbrace -1, 1 \rbrace \ ,
\end{equation} 
where the couplings $J_{ij}$ and the fields $h_i$ are i.i.d. random variables chosen from a normal distribution $\mathcal{N}(0,1)$. 
We also considered the same model with a Boolean spin representation:
\begin{equation} \label{eqH:Boolean}
H=  \sum_i \left( 2 \sum_{ j  \in \partial i} J_{ij} -2 h_i \right) \sigma_i - 4 \sum_{(ij)} {J_{ij}} \sigma_i \sigma_j - \sum_{(ij)} J_{ij}  + \sum_i h_i \ , 
\hskip30pt \sigma_i=\lbrace 0,1 \rbrace \ .
\end{equation}
Note that in this case no exact solution is available. Thus we compared our expansion with the result of
Monte Carlo simulations, which provide an alternative way to access to the magnetizations and correlations of the system. 
With Monte Carlo,
one can also compute 
the free energy $W$ by thermodynamic integration, using the relation:
 \begin{equation}
 \frac{\partial W}{\partial \beta}= - \langle H \rangle_\beta \qquad \mbox{so} \qquad  W(\beta) =  W(0) - \int_{0}^{\beta} \langle H \rangle_{\beta'}  d \beta' \ ,
 \label{thermo_int}
 \end{equation}
with $W(0)=N \log(2)$ (where $N$ is the number of spins of the system).
We computed the average energy $\langle H \rangle $ with a 
Monte Carlo-Metropolis algorithm~\cite{d} and performed numerically the integration using the Simpson's rule.
The computational time is of order $\mathcal{O}(MC \cdot INT)$ with $MC$ being the number  
of Monte Carlo steps (including the thermalization phase) and $INT$ the number of
intervals needed to perform the numerical integration.

In Figure \ref{EAG} we plot the results for a system with $10 \times 10$ spins.
We observe that for the free energy both the Ising and Boolean expansions
converge reasonably well, but the Ising expansion is faster. 
On the contrary we note that in the case $\beta=0.7$, even if the free energy seems to converge, this is not the case for the connected correlations.

\subsection{Sherrington-Kirkpatrick model}

In this section, we report the results of our algorithm when applied to disordered Ising models on a complete graph.
We consider the case with $N= 100$ spins, i.i.d. random external fields $h_i \sim {\cal N}(0,1)$ and scaled couplings $J_{ij} \sim {\cal N}(0,1)$
(Sherrington-Kirkpatrick or SK model)~\cite{mpv}:
\begin{equation}
H = -\sum_{i<j} \frac{J_{ij}}{\sqrt{N}} s_i s_j - \sum_i h_i s_i  \ , \qquad s_i= \pm 1 \ .
\end{equation}
An exact solution of this model is known in the thermodynamic limit $N\to\infty$~\cite{mpv}. At the ``replica symmetric'' level, the solution is
\begin{equation}\label{free_annealed_SK}
\begin{split}
\frac{W}{N} &= -\beta f = \frac{(1-q)^2}{4 T^2} + \int {\cal D}z \log\{2 \cosh [\beta (\sqrt{q} z + H) ]  \}  \ , \\
q &= \int {\cal D}z \tanh^2 [\beta (\sqrt{q} z + H) ] \ .
\end{split}\end{equation}
Here, ${\cal D}z =  dz \frac{e^{-z^2/2}}{\sqrt{2\pi}}$ is a Gaussian integration measure.

\begin{figure}[H]
\centering
\includegraphics[width=.9\textwidth]{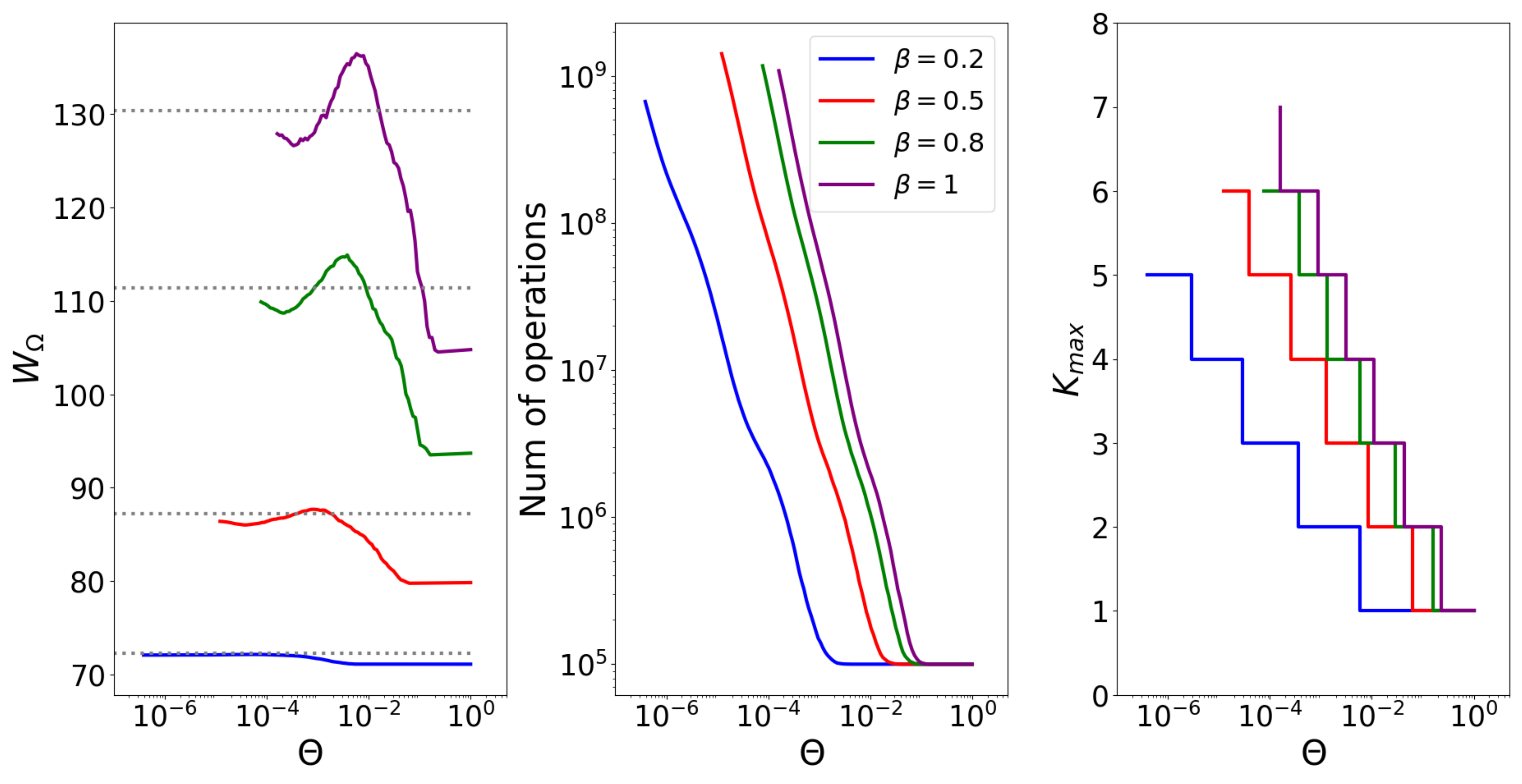}
\caption{ Performance of the ACE for the SK model, in the Ising spin representation.
The black dashed line is the exact result Eq.~(\ref{free_annealed_SK}).}
\label{fig:SK}
\end{figure}

In this case, the algorithm seems to converge towards the correct result, although with large oscillations, see Figure \ref{fig:SK}.
Large oscilllations are due to the fact that the model is fully connected so at each cluster size there is a large number of relevant clusters
that need to be evaluated. Therefore, although $K_{max}$ remains relatively small ($K_{max} \leq 6$), the number of operations
becomes quickly very large because of the large number of clusters, preventing an efficient evaluation of the free energy. We conclude
that the algorithm is not very efficient for densely connected interaction graphs. Note that we did not explore the low-temperature
spin glass phase below the de Almeida-Thouless line~\cite{mpv}.

\subsection{Erd\"os-R\'enyi random graph model}

In this section we report the results of our algorithm when applied to disordered Ising models on diluted random graphs
with finite connectivity.
The random networks are generated from the Erd\"os-R\'enyi ensemble, where $M=\frac{c}{2}N$ edges are drawn, uniformly and at random, between $N$ points with $c$ the average degree of a vertex. On the selected bonds $(ij)$  the couplings $J_{ij}$ are chosen from $\mathcal{N}(0,1)$. All other couplings and the fields $h_i$ are set to zero (Viana-Bray model)~\cite{hw,mm}:
\begin{equation}
H = -\sum_{(ij)} J_{ij} s_i s_j \ , \qquad s_i=\lbrace -1 , 1\rbrace \ .
\end{equation}
In the paramagnetic phase, the solution can be obtained via the cavity method~\cite{mm} and reads
\begin{equation} \label{ER_sol}
\frac{W}{N}=\log(2) + \frac{c}{2} \overline{ \log(\cosh \beta J_{ij}) }  \ .
\end{equation} 
This solution holds if $c \, \overline{\tanh^2(\beta J_{ij})}\leq 1$. For example,
for $c=5$, the paramagnetic phase is stable for $\beta < \beta_c = 0.5531\cdots$.
In the spin glass phase, no exact solution is available but $W$ can be estimated via Monte Carlo for small enough $N$.

\begin{figure}[t]
 \centering
\includegraphics[width=.9\textwidth]{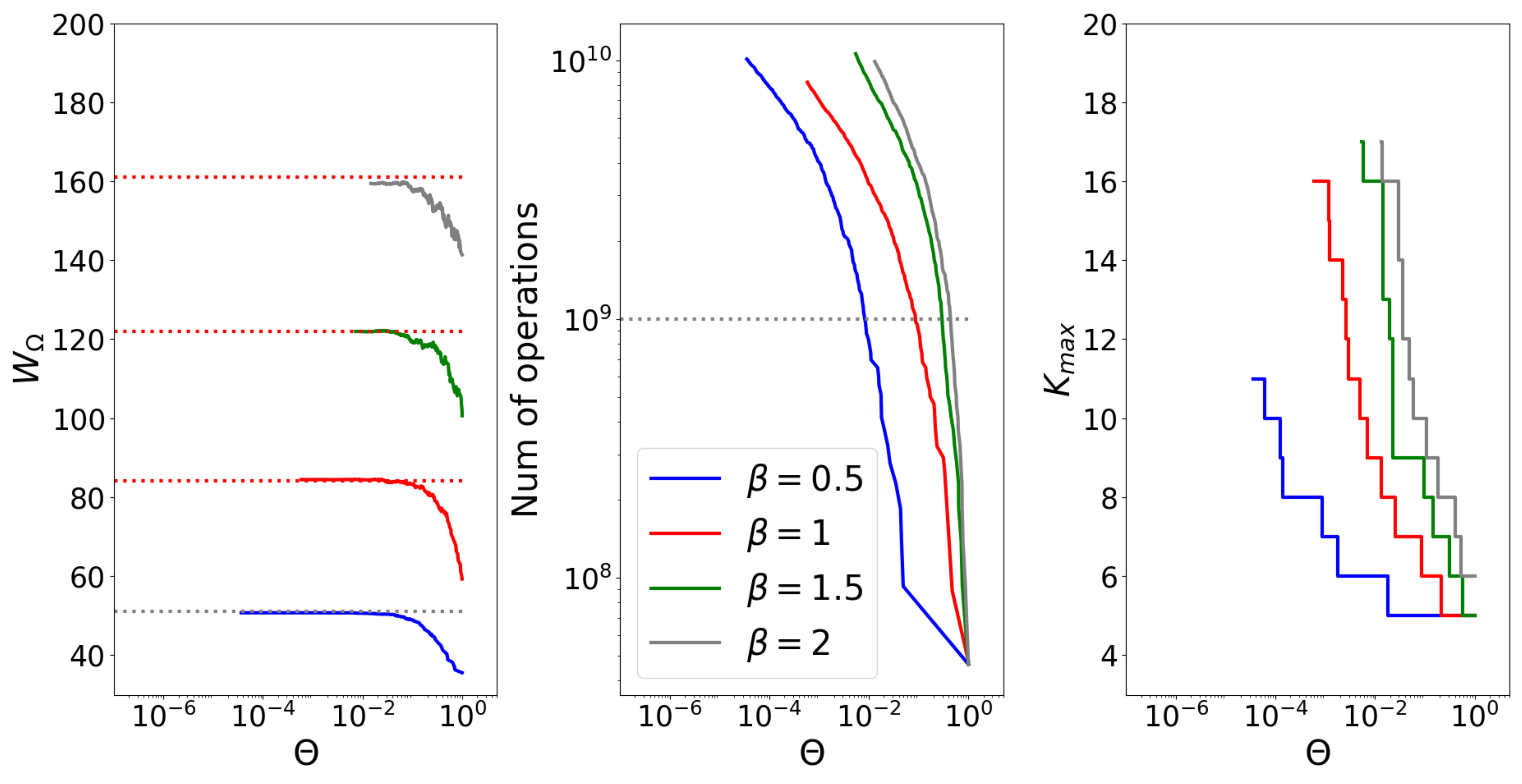}
\caption{Performances of the ACE in the Erd\"os-R\'enyi model with $N=50$, $c=5$, without magnetic fields and setting $K_{min} = 4$, in the Ising representation.
The dashed black line is the cavity solution Eq.~\eqref{ER_sol}, while the red dashed lines are the result of Monte Carlo simulations in the spin glass phase.}
\label{erdos2}
\end{figure}

When applying the ACE algorithm to this model we noted that the cluster expansion
is truncated at  small $K_{max}$. The reason is that on this particular graph, there are few closed polygons 
which are the only ones with $\Delta W_\Gamma \neq 0$ according to Eq.~\eqref{eq:Zpolygon}.
All the others clusters have $\Delta W_\Gamma=0$ and are thus discarded, similarly to the one-dimensional model.
To avoid this problem we set a a minimal cluster length $K_{min}=4$, meaning that all clusters with size less than or equal to 4 are considered significant and used in the ACE algorithm.
This procedure allows to consider longer cluster thus improving the convergence of the expansion (see Figure \ref{erdos2}).
Note that setting $K_{min}=4$ we can construct clusters of length 5, thus the minimal number of operation is $\mathcal{O}{{50}\choose{5}} \sim 10^7$ and the size of the largest significant clusters $K_{max}$ is at least 5 for each $\Theta$. 

\begin{figure}[t]
 \centering
\includegraphics[width=1\textwidth]{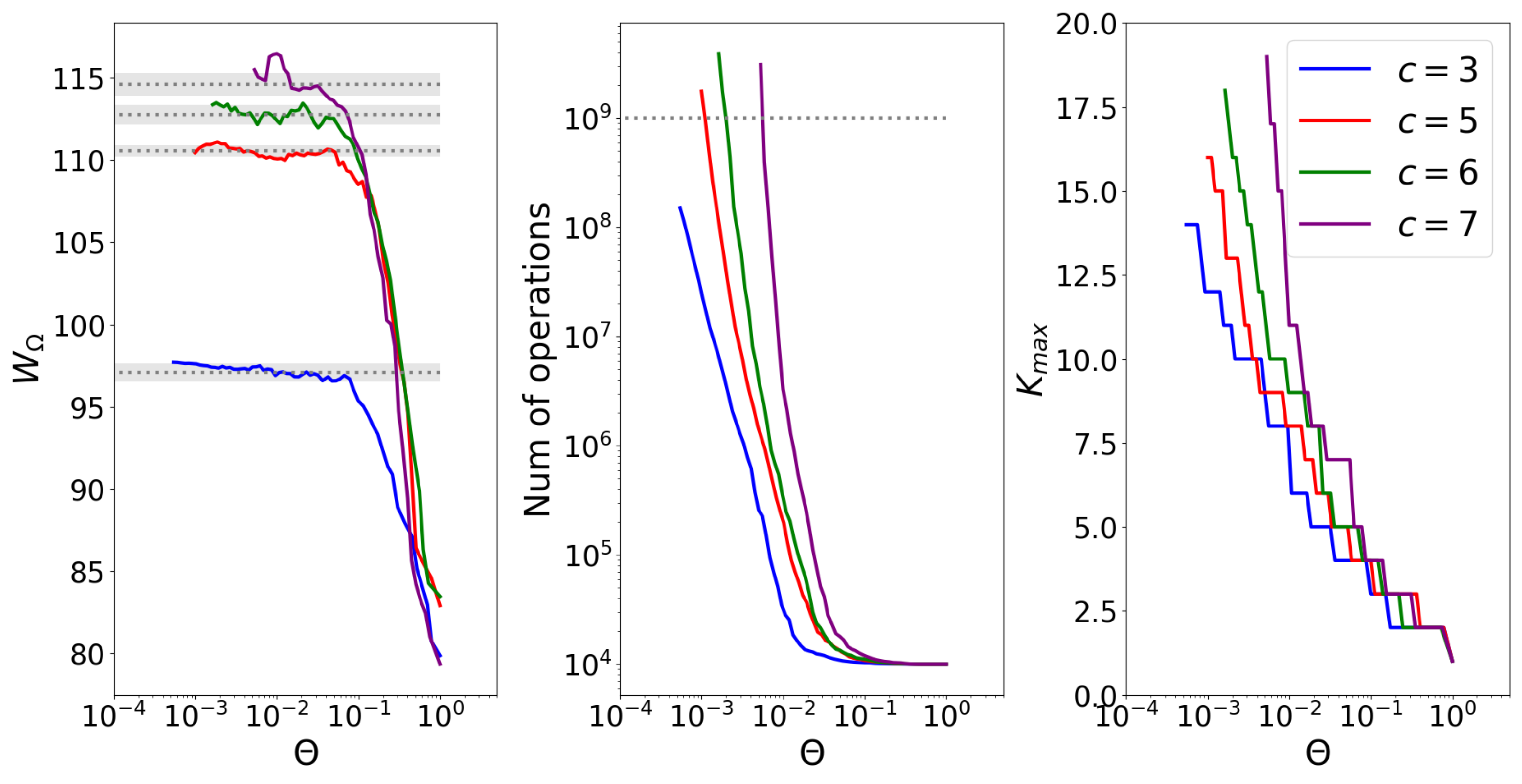}
\caption{
Performances of the ACE in the Erd\"os-R\'enyi model with random magnetic fields, at $T=2$ and several $c$, in the Ising representation.
In the left panel, the dotted lines are the average of 10 Monte Carlo simulation (with the shaded error bar being the standard deviation), 
each of them requiring a computational time $\mathcal{O}(10^9)$:
 we  set a total of $MC=10^7$ Monte Carlo steps and $INT=100$ (the number of intervals to perform the numerical integration).
 The corresponding total number of iterations is indicated by a dotted line in the middle panel.
 }
\label{erdos_campi}
\end{figure}

As in the previous cases, we also considered the case where random fields $h_i \sim \mathcal{N}(0,1)$ 
 are added to the Hamiltonian. In Figure~\ref{erdos_campi} we report results for this model with $N=100$, temperature $T=2$,
 and several connectivities. In this case the system is paramagnetic, but due to the presence of external fields the cavity solution
 is more complicated~\cite{mm}.  
We thus compare the ACE with a Monte Carlo simulation, finding good agreement.

\subsection{Application on biological data}

\begin{figure}[t]
	\centering
	\includegraphics[width=.9\textwidth]{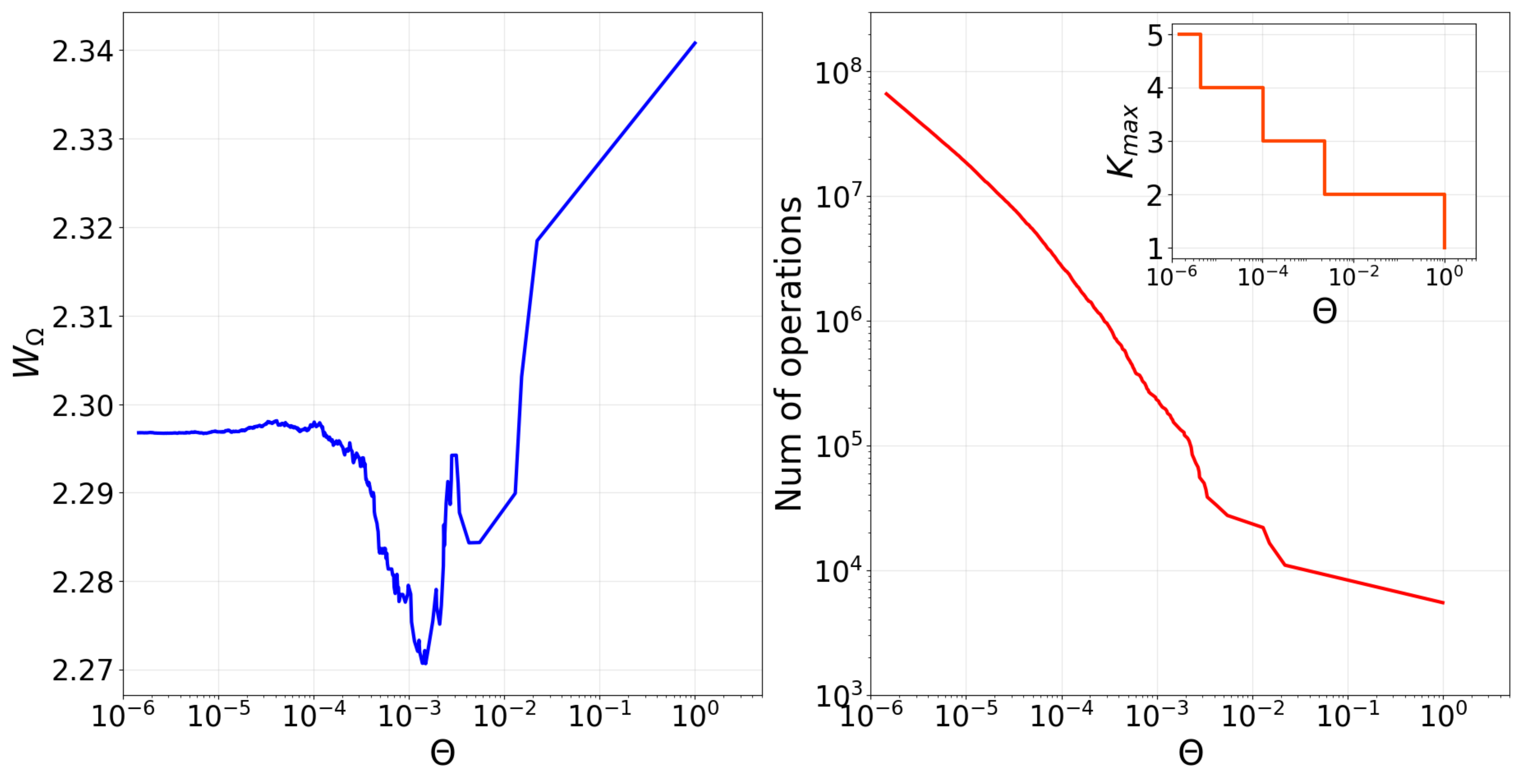}
	\includegraphics[width=.9\textwidth]{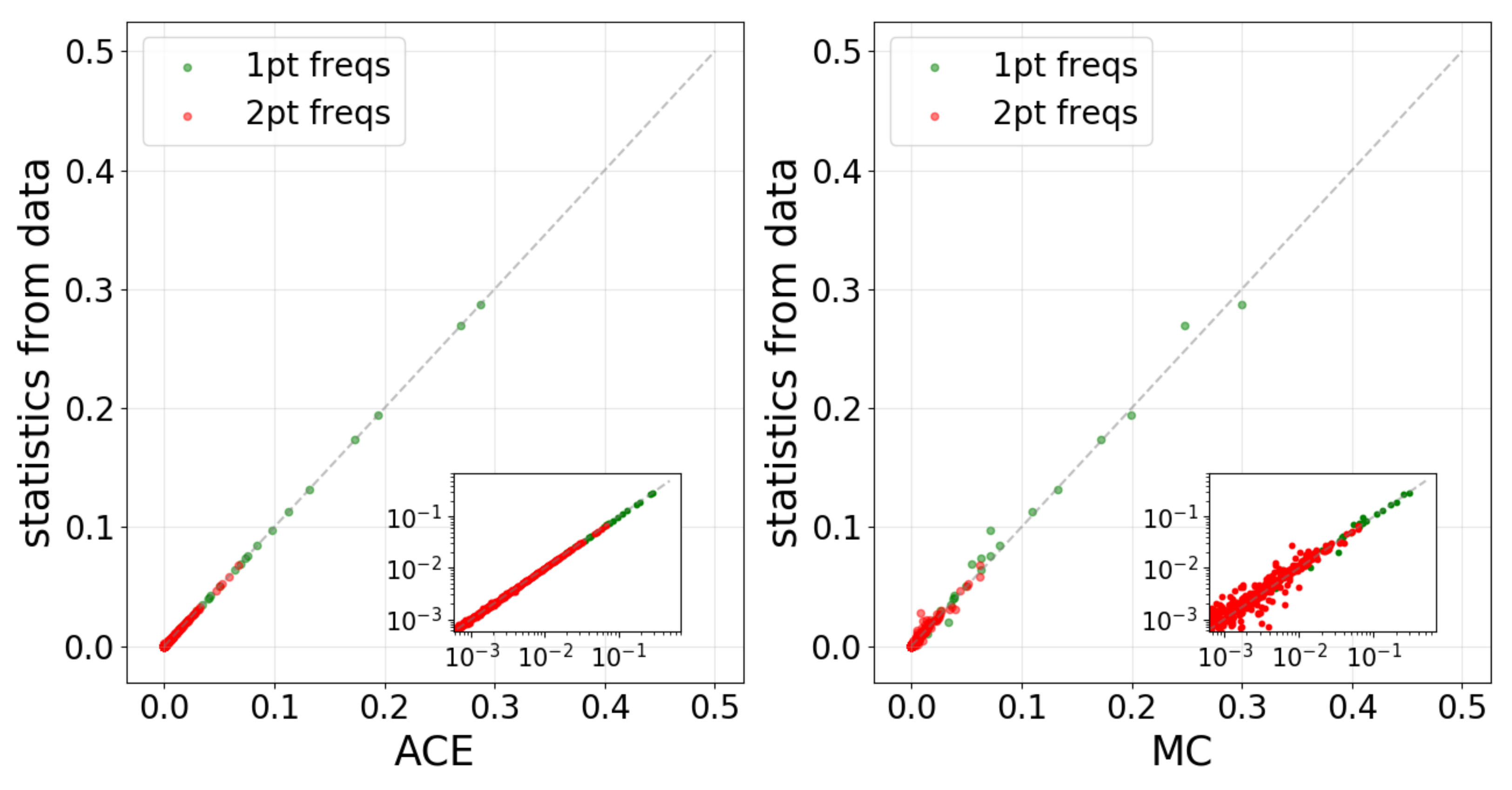}
	\caption{Application of the ACE to a model of interacting neurons inferred from data, in the Boolean representation.
	In the top row, the free energy and the number of operations of the ACE algorithm. Note that the free energy converges in less than $10^8$ operations.
		In the bottom right plot, the one- and two-points frequencies computed with our algorithm at $\Theta \simeq 10^{-6}$ compared with the empirical ones.
	In the bottom left plot, the Monte Carlo results which requires computational time $\mathcal{O}(10^8)$. }
	\label{biodata}
\end{figure}

We also applied our algorithm on data coming from the prefrontex cortex.
A population of $N=37$ neurons is recorded and their interactions are represented by a fully connected Boolean spin Hamiltonian, Eq.~\eqref{eqH:Boolean},
inferred using the inverse ACE algorithm~\cite{Barton-Cocco}. Note that the inferred fields are of the order of $h \sim -4$,
while the inferred couplings are of the order of $J \sim 0.1$.
Here we apply our algorithm on the inferred model. 
We note that in less than $10^8$ operations, the free energy converges and the one- and two-points (non-connected) frequencies  
$m_i = \langle \sigma_i \rangle$, $f_{ij} = \langle \sigma_i \sigma_j \rangle $ are well correlated with those coming from the data.
On the contrary, a Monte Carlo sampling with the same number of operations is less correlated with the statistics from the data (see Figure \ref{biodata}).
Hence, in this case, the ACE algorithm appears to be faster and more reliable than the Monte Carlo.

\section{Conclusions}
\label{sec:conclusion}

In this paper we have presented an Adaptive Cluster Expansion (ACE) algorithm for the direct Ising problem, and
we discussed its application to some benchmark systems, both ordered and disordered, and with different topologies of the interaction network. 
Although each case has its own specificities, we can draw
some general conclusions. 
In ordered cases, the simple recursive construction rule of the ACE algorithm can get stuck because of the vanishing of cluster contributions 
due to symmetries, but it is easy to adapt the construction rule to take symmetries into account. 
Indeed, one can classify the relevant clusters and find numerically their contributions, which can be summed 
with the correct multiplicity to obtain an accurate estimation of the free energy. 
An example has been given in section~\ref{sec:examples-ord},
for the two-dimensional ferromagnetic Ising model, in which only one-spin, two-spin and closed-loop clusters contribute. Another example is given by the Erd\"os-R\'enyi model
without external field, in which all three-spin cluster contributions vanish and we had to set a $K_{min}=4$ in the construction rule.
The simple, recursive, construction rule is particularly adapted
to treat cases with disorder where there are no special symmetries that
make some clusters vanishing (such as the inversion symmetry). Also, diluted systems are more efficiently treated by the algorithm. On the contrary,
the ACE scheme runs into difficulties for densely connected systems.

Because of the locality of the cluster expansion, the choice of variable representation, e.g. Boolean versus Ising,
turns out to be important for the convergence of the expansion.
As we have shown in the one-dimensional Ising model, when the representation is changed from Ising to Boolean variables,  
even in absence of external fields in the original Ising model, non zero fields arise from the change to Boolean variables.
As a consequence,
while the expansion in Ising variables is around independent paramagnetic spins, the expansion for Boolean variables is around strongly magnetized spins. 
While Ising variables seem more appropriate for the free energy expansion 
in the particular case of zero field and total inversion symmetry, this is not the case for the correlation functions which are set to zero for spins at distance larger than two
in the Ising representation, where 
clusters larger than two have zero free energy contribution.
The quality of the approximation for the free energy at small cluster size depends on the representation, 
while the cluster contributions decay is controlled in both cases by the correlation length of the model.
To calculate correlations between two sites we need clusters which include these sites, which implies that depending on where the expansion stops,
some correlations are set to zero, and an error is made which depends on the correlation length of the system.  
In these cases, one can add a small magnetic field and take a numerical derivative to obtain the correlations.
In summary, the Boolean representation is more adapted for systems with strong
magnetic fields, such as the neuronal system studied in Figure~\ref{biodata}, while the Ising representation is more adapted for systems with weak magnetic fields.
Note that the Ising and Boolean representations are only two special cases in a continuum of possible representations where the spin variables take two generic values $a$ and $b$. Given a model, one could, at least in theory, identify an ``optimal'' representation, meaning two values $a$ and $b$ such that the ACE algorithm 
converges most quickly. The performance of the algorithm depends in a non-trivial way on $a$ and $b$, because to determine the optimal representation one should a-priori compute the Hamiltonian and cluster symmetries. Here we did not explore the problem further, but see Ref.~\cite{rizzato} for a discussion of this point in the context of the inverse problem.

In the case of disordered systems, for which no exact solution is available, and cluster contributions are heterogeneous,
we compared our ACE algorithm to the standard Monte Carlo method.
While Monte Carlo simulation is a well-established and powerful tool in
statistical mechanical calculations, it is well known that the calculation of free
energies is a difficult problem for this algorithm, as it requires a thermodynamic integration, e.g. as in Eq.~(\ref{thermo_int}).
This method requires the computation of the mean energy of a system at different
temperatures and the computational time is of order $\mathcal{O}(MC \cdot
INT)$ where $MC$ is the number of Monte Carlo steps  (including the
thermalization phase) and $INT$ the number of
intervals needed to perform the numerical integration correctly. These two parameters, which
strongly influence the quality of the simulation, depend on
the system size and on the system Hamiltonian, and optimizing their value is not straightforward because of the difficulty of 
checking the equilibration of the Monte Carlo.
On the contrary the ACE algorithm depends on a single parameter $\Theta$, the cluster expansion threshold, which can be lowered until convergence, 
if the number of operations does not exceed the available computational time. Convergence of the observables can thus be monitored directly.
We found that its performances are comparable, or in some cases better, than
Monte Carlo. We therefore believe that the ACE could be a useful tool for the free energy calculation
(possibly adapting the cluster summation rule to the symmetries of the system),
in cases where the network is not too dense, no
symmetries are present, and disorder is strong enough to slow down the MC equilibration considerably.
Note that in this paper we did not test more advanced MC schemes, that might be more efficient than the ACE in some cases,
especially on large systems.
We believe that no general conclusion can be drawn, and one should test in each case which is the best performing algorithm.

Note that the ACE algorithm can be easily adapted for spin variables taking more than two values. Indeed, the expansion in Eq.~(\ref{clust_approx}) is independent of the Hamiltonian of the system and is valid for any arbitrary function $W$. 
However, to this end, one should modify the routines Algorithm 2 and 3, and this extension can be computationally expensive. Indeed, the ACE algorithm performs an exact computation of $W$ for each cluster; it requires a computational time $\mathcal{O}(2^k)$ (with $k$ the size of the cluster) which becomes $\mathcal{O}(q^k)$ when considering spin variables taking $q$ values.
In our numerical experiments (with $q=2$) we checked that we can construct maximum clusters containing 20 spins. Therefore, for a $q$-state spin variable, the maximum size of the cluster is $K_{max}= {20\log 2}/{\log q}$. As a reference, we can consider the case of $q=21$ which is particularly relevant since it is the model used for the prediction of structural contacts from protein sequences (the inverse ACE algorithm was originally developed for this purpose \cite{Monasson-Cocco}).
In this case $K_{max} \sim 5$, and, in some situations, the direct ACE algorithm could fail to reach convergence with such small clusters. Yet, ``compression'' tricks can be
implemented to reduce the effective number of states~\cite{rizzato}.
In principle, the ACE algorithm could also be adapted to more general spin systems such as the Heisenberg model \cite{rew1_EPJB_c,baxter}, classical XY model \cite{baxter} or Blume-Capel model \cite{blume1971ising}.
Nevertheless, the computation of the cluster contribution $\Delta W$ with continuous spin variables can become more problematic and we did not explore further the possibility.

The ACE algorithm could be improved in several ways. One could for example perform the expansion in
Eq.(\ref{clust_approx}) in $W_{\Omega}(\mathcal{J}) - W_{0}$, where $W_{0}$ is a reference value obtained, for example, via
mean field~\cite{Monasson-Cocco2}. 
This procedure might be helpful for fully connected systems, for which we have seen that the adaptive cluster expansion does not converge well. 

It is important to notice that the  cluster expansion can be applied to any  function, not only the log  partition function of an Ising  model  with pairwise 
couplings on different interaction graph, as discussed in this paper. 
An example is the cross-entropy, as originally implemented for the inverse problem~\cite{Monasson-Cocco}.
For instance, the cluster expansion can also be used to study the stationary states of a dynamical  systems, by considering spatio-temporal evolutions of configurations. 
Dynamical cluster expansions have been recently built to study for example the spreading of epidemic diseases~\cite{petermann04,barthel05,Pelizzola2013}.
The adaptive cluster expansion can also be adapted to  include temporal correlations~\cite{obermayer14}. 

\section*{Acknowledgements}

The code we implemented for the direct Ising problem is available online at 
\url{https://github.com/GiancarloCroce/directACE}.
It is based on code used for the inverse problem in Ref.~\cite{Monasson-Cocco2}, 
\url{https://github.com/johnbarton/ACE}.
We thank John Barton and Remi Monasson for useful discussions.

\bigskip

\textbf{References}
\bibliography{references_v1} 
\bibliographystyle{unsrt}

\end{document}